# Practical Locally Private Heavy Hitters[*]

Raef Bassily[†]    Kobbi Nissim[‡]    Uri Stemmer[§]    Abhradeep Thakurta[¶]

July 16, 2017

**Abstract**

We present new practical local differentially private heavy hitters algorithms achieving optimal or near-optimal worst-case error and running time – TreeHist and Bitstogram. In both algorithms, server running time is $\tilde{O}(n)$ and user running time is $\tilde{O}(1)$, hence improving on the prior state-of-the-art result of Bassily and Smith [STOC 2015] requiring $O(n^{5/2})$ server time and $O(n^{3/2})$ user time. With a typically large number of participants in local algorithms ($n$ in the millions), this reduction in time complexity, in particular at the user side, is crucial for making locally private heavy hitters algorithms usable in practice. We implemented Algorithm TreeHist to verify our theoretical analysis and compared its performance with the performance of Google's RAPPOR code.

## 1  Introduction

We revisit the problem of computing heavy hitters with local differential privacy. Such computations have already been implemented to provide organizations with valuable information about their user base while providing users with the strong the strong guarantee that their privacy would be preserved even if the organization is subpoenaed for the entire information seen during an execution. Two prominent examples are Google's use of RAPPOR in the Chrome browser [11] and Apple's use of differential privacy in iOS-10 [19]. These tools are used for learning new words typed by users and identifying frequently used emojis and frequently accessed websites.

**Differential privacy in the local model.** Differential privacy [10] provides a framework for rigorously analyzing privacy risk and hence can help organization mitigate users' privacy concerns as it ensures that what is learned about any individual user would be (almost) the same whether the user's information is used as input to an analysis or not.

Differentially private algorithms work in two main modalities: curator and local. The *curator model* assumes a trusted centralized curator that collects all the personal information and then analyzes it. In contrast, the *local model* does not involve a central repository. Instead, each piece of personal information is randomized by its provider to protect privacy even if all information provided to the analysis is revealed. Holding a central repository of personal information can become a liability to organizations in face of security breaches, employee misconduct, subpoenas, etc. This makes locally private computations attractive for implementation. Indeed in the last few years Google and Apple have announced successful deployments of locally private analyses [11, 19].

**Challenges of the local model.** A disadvantage of the local model is that it requires introducing noise at a significantly higher level than what is required in the curator model. Furthermore, some tasks tasks which are possible in the curator model are impossible in the local model [10, 16, 8]. To see the effect of noise, consider estimating the number of HIV positives in a given population of $n$ participants. In the curated model, it suffices to add Laplace noise of magnitude $O_\epsilon(1)$, i.e., independent of $n$ [10]. In contrast, a lowerbound of $\Omega_\epsilon(\sqrt{n})$ is known for the local model [8]. A higher noise level implies that the number of participants $n$ needs to be large (maybe in the millions for a reasonable choice of $\epsilon$). An important consequence is that, to be practical, locally private algorithms must exhibit low time, space, and communication complexity, especially at the user side. This is the problem addressed in our work.

---

[*]Research by K.N. and U.S. supported by NSF grant No. 1565387.

[†]Center for Information Theory and Applications and Department of Computer Science & Engineering, University of California San Diego. rbbassily@gmail.com

[‡]Department of Computer Science, Georgetown University *and* Center for Research on Computation and Society (CRCS), Harvard University. kobbi.nissim@georgetown.edu.

[§]Center for Research on Computation and Society (CRCS), Harvard University. u@uri.co.il

[¶]Department of Computer Science, University of California Santa Cruz. aguhatha@ucsc.edu.

**Heavy hitters and histograms in the local model.** Assume each of $n$ users holds an element $x_i$ taken from a domain of size $d$. A histogram of this data lists (an estimate of) the duplicity of each domain element. When $d$ is large, a succinct representation of the histogram is desired either in form of a *frequency oracle* – a data structure allowing to approximate the duplicity of any domain element – or *heavy hitters* – listing the duplicity of most frequent domain elements, implicitly considering the duplicity of other domain elements as zero. The problem of computing histograms with differential privacy has attracted significant attention both in the curator model [10, 5, 7] and the local model [15, 11, 4]. Of relevance is the work in [17].

We briefly report on the state of the art heavy hitters algorithms of Bassily and Smith [4] and Thakurta et al. [19], which are most relevant for the current work. Bassily and Smith provide matching lower and upper bounds of $\Theta(\sqrt{n \log(d)}/\epsilon)$ on the worst-case error of local heavy hitters algorithms. Their local algorithm exhibits optimal communication but a rather high time complexity: Server running time is $O(n^{5/2})$ and, crucially, user running time is $O(n^{3/2})$ – complexity that severely hampers the practicality of this algorithm[1]. The construction by Thakurta et al. is a heuristic with no bounds on server running time and accuracy.[2] User computation time is $\tilde{O}(1)$, a significant improvement over [4].

**Our contributions.** The focus of this work is on the design of locally private heavy hitters algorithms with near optimal error, keeping time, space, and communication complexity minimal. We provide two new constructions of heavy hitters algorithms TreeHist and Bitstogram. These algorithms achieve similar performance but apply different techniques. We implemented Algorithm TreeHist and provide measurements in comparison with RAPPOR [11] (the only currently available implementation for local histograms). Our measurements are performed with a setting that is favorable to RAPPOR (i.e., a small input domain), yet they indicate that Algorithm TreeHist performs better than RAPPOR in terms of noise level.

Table 1 details various performance parameters of algorithms TreeHist and Bitstogram, and the reader can check in the table that these are similar up to small factors which we ignore for the rest of this paragraph. Comparing with [4], we improve time complexity both at the server (reduced from $O(n^{5/2})$ to $\tilde{O}(n)$) and at the user (reduced from $O(n^{3/2})$ to $O(\max(\log n, \log d)^2)$). Comparing with [19], we get provable bounds on the server running time and worst-case error. Note that Algorithm Bitstogram achieves optimal worst-case error whereas Algorithm TreeHist is almost optimal, by a factor of $\sqrt{\log(n)}$.

| Performance metric | TreeHist | Bitstogram |
|---|---|---|
| Server time (modular multiplications) | $\tilde{O}(n)$ | $\tilde{O}(n)$ |
| User time (modular multiplications) | $O\left(\max(\log n, \log d)^2\right)$ | $O\left(\max(\log n, \log d)^2\right)$ |
| Server processing memory | $\tilde{O}(\sqrt{n})$ | $\tilde{O}(\sqrt{n})$ |
| User memory | $O(\max(\log d, \log n))$ | $O(\max(\log d, \log n))$ |
| Communication/user | $O(1)$ | $O(1)$ |
| Worst-case Error | $O\left(\sqrt{n \log(n) \log(d)}\right)$ | $O\left(\sqrt{n \log(d)}\right)$ |

Table 1: Performance of our protocols. Dependency on the privacy parameter $\epsilon$ and failure probability $\beta$ is omitted.

**Elements of the constructions.** Main details of our constructions are presented in sections 3.1 and 3.2. These are complemented with the detailed descriptions and analyses in the appendix. Both our algorithms make use of frequency oracles – data structures that allow estimating various counts.

Algorithm TreeHist identifies heavy-hitters and estimates their frequencies by scanning the levels of a binary prefix tree whose leaves correspond to dictionary items. The recovery of the heavy hitters is in a bit-by-bit manner. As the algorithm progresses down the tree it prunes all the nodes that *cannot* be prefixes of heavy hitters, hence leaving $\tilde{O}(\sqrt{n})$ nodes in every depth. This is done by making queries to a frequency oracle. Once the algorithm reaches the final level of the tree it identifies the list of heavy hitters. It then invokes the frequency oracle once more on those particular items to obtain more accurate estimates for their frequencies.

---

[1] If pubic randomness or efficient Private Information Retrieval is assumed, then the user running time in [4] can be improved to $O(n)$.
[2] The underlying construction in [19] is of a frequency oracle.

Algorithm Bitstogram hashes the input domain into a domain of size roughly $\sqrt{n}$. The observation behind this algorithm is that if a heavy hitter $x$ does not collide with other heavy hitters then $(h(x), x_i)$ would have a significantly higher count than $(h(x), \neg x_i)$ where $x_i$ is the $i$th bit of $x$. This allows recovering all bits of $x$ in parallel given an appropriate frequency oracle.

# 2 Background and Preliminaries

## 2.1 Definitions and Notation

**Dictionary and user items:** Let $\mathcal{V} = [d]$. We consider a set of $n$ users, where each user $i \in [n]$ holds an item $v_i \in \mathcal{V}$. We will use $v_i$ to refer to the binary representation of $v_i$ when it is clear from the context.

**Item frequency (duplicity):** For each item $v \in \mathcal{V}$, we define the frequency $f(v)$ of $v$ as the number of users holding $v$, namely,
$$f(v) \triangleq |\{i \in [n] : v_i = v\}|.$$

**Frequency oracle:** A frequency oracle is a data structure together with an algorithm that, for any given $v \in \mathcal{V}$, allows computing an estimate $\hat{f}(v)$ of the frequency $f(v)$.

**Heavy hitters (succinct histogram):** A succinct histogram is a data structure that provides a (short) list of items $(\hat{v}_1, ..., \hat{v}_k)$, called the *heavy hitters*, together with estimates for their frequencies $(\hat{f}(\hat{v}_j) : j \in [k])$. The frequencies of the items not in the list are implicitly estimated as $\hat{f}(v) = 0$. We measure the error in a succinct histogram by the $\ell_\infty$ distance between the estimated and true frequencies, $\max_{v \in [d]} \left| \hat{f}(v) - f(v) \right|$. We will also consider the maximum error restricted to the items in the list $(\hat{v}_1, ..., \hat{v}_k)$, that is, $\max_{j \in [k]} \left| \hat{f}(\hat{v}_j) - f(\hat{v}_j) \right|$. If a succinct histogram aims to provide $\ell_\infty$ error $\eta$, the list does not need to contain more than $O(1/\eta)$ items since items with estimated frequencies below $\eta$ may be omitted from the list.

## 2.2 Local Differential Privacy

In the local model, an algorithm $\mathcal{A}: \mathcal{V} \to \mathcal{Z}$ accesses the database $\mathbf{v} = (v_1, \ldots, v_n) \in \mathcal{V}^n$ only via an oracle that, given index $i \in [n]$, runs a randomized algorithm (local randomizer) $\mathcal{R}: \mathcal{V} \to \tilde{\mathcal{Z}}$ on input $v_i$ and returns $\mathcal{R}(v_i)$ to $\mathcal{A}$.

**Definition 2.1** (Local differential privacy [10, 12, 16]). *An algorithm satisfies $\epsilon$-local differential privacy (LDP) if it accesses the database $\mathbf{v} = (v_1, \ldots, v_n) \in \mathcal{V}^n$ only via invocations of a local randomizer $\mathcal{R}$ and if for all $i \in [n]$, if $\mathcal{R}^{(1)}, \ldots, \mathcal{R}^{(k)}$ denote the algorithm's invocations of $\mathcal{R}$ on the data sample $v_i$, then the algorithm $\mathcal{A}(\cdot) \triangleq \left( \mathcal{R}^{(1)}(\cdot), \mathcal{R}^{(2)}(\cdot), \ldots, \mathcal{R}^{(k)}(\cdot) \right)$ is $\epsilon$-differentially private. That is, if for any pair of inputs $\mathbf{v}, \mathbf{v}'$ that differ on a single input, and for all $\mathcal{S} \subseteq \mathsf{Range}(\mathcal{A})$,*
$$\Pr[\mathcal{A}(v) \in \mathcal{S}] \leq e^\epsilon \cdot \Pr[\mathcal{A}(v') \in \mathcal{S}].$$

## 2.3 Count Sketch and Hadamard Transform

Count sketch [9] together with Hadamard transform form the basis of our differentially private construction outlined in Section 3.1 and discussed in detail in Section 5.

**Count sketch [9]** is a sketching algorithm for finding frequent elements in a data stream. Let $\mathcal{V} = [d]$ be a domain of data elements, and $\mathbf{v} = (v_1, \ldots, v_n)$ be a stream of data elements. Count sketch ensures that for any given $v \in \mathcal{V}$, using a data structure of size $m = O\left( \frac{n \log(1/\beta)}{k} \right)$ one can ensure that with probability at least $1 - \beta$ the estimated frequency of $v$ is within $k$ of the true frequency, and the estimate is *unbiased*. The algorithm works as follows: First, pick $t$ pairs of hash functions $(h_i : \mathcal{V} \to [m], g_i : \mathcal{V} \to \{-1, 1\})$, and set a matrix $\mathbf{M} = \{0\}^{t \times m}$. Second, with every data sample $v_i$, populate the matrix as follows: $\forall j \in [t], \mathbf{M}[j, h_j(v_i)] \leftarrow \mathbf{M}[j, h_j(v_i)] + g_j(v_i)$. Finally, to estimate the frequency of an element $v \in \mathcal{V}$, compute median $\{\mathbf{M}[1, h_1(v)] \cdot g_1(v), \cdots, \mathbf{M}[t, h_t(v)] \cdot g_t(v)\}$.

**Hadamard transform:** We use Hadamard transform followed by sampling in order to compress our data transmission from the client to the server and to reduce the space requirements of our protocol. Hadamard transform of a vector $w \in \mathbb{R}^m$ is obtained via multiplying with the Hadamard transform matrix $\mathbf{H}_m \in \{-\frac{1}{\sqrt{m}}, \frac{1}{\sqrt{m}}\}^{m \times m}$ defined recursively as $\mathbf{H}_m = \frac{1}{\sqrt{2}} \begin{bmatrix} \mathbf{H}_{m/2} & \mathbf{H}_{m/2} \\ \mathbf{H}_{m/2} & -\mathbf{H}_{m/2} \end{bmatrix}$, and $\mathbf{H}_1 = [1]$. Two main properties of Hadamard transform we use are: i) it is a dense basis transformation, i.e. the columns of the Hadamard matrix form a basis, and each entry of $\sqrt{m} \cdot \mathbf{H}_m$ is in $\{-1, 1\}$, and ii) any entry $(i, j)$ in $\sqrt{m} \cdot \mathbf{H}_m$ can be computed in $O(\log m)$ time.

## 2.4 Error correction codes

We will use error correction codes in order to reduce the error of (some of) our constructions outlined in Section 3.2 and discussed in detail in Section 6.

**Definition 2.2.** *A binary $(n, k)$-code is a pair of mappings* $(\mathrm{Enc}, \mathrm{Dec})$ *where* $\mathrm{Enc} : \{0,1\}^k \to \{0,1\}^n$*, and* $\mathrm{Dec} : \{0,1\}^n \to \{0,1\}^k$*. The code is $\zeta$-decodable if for every $x \in \{0,1\}^k$ and every $y \in \{0,1\}^n$ whose Hamming distance to $\mathrm{Enc}(x)$ is at most $\zeta n$ we have that $\mathrm{Dec}(y) = x$.*

For any constant $0 < \zeta < 1/4$, there is a construction of a $\zeta$-decodable $(n, k)$-code, where $n = O(k)$, and furthermore, $\mathrm{Enc}$ and $\mathrm{Dec}$ run in time $O(n)$. See, e.g., [14].

## 2.5 Tools from Probability

### 2.5.1 $k$-wise Independence

We will use the following tail bound on sums of $k$-wise independent random variables.

**Lemma 2.3** ([6]). *Let $\lambda \geq 6$ be an even integer. Suppose $X_1, \cdots, X_n$ are $k$-wise independent random variables taking values in $[0,1]$. Let $X = X_1 + \cdots + X_n$ and $\mu = \mathbb{E}[X]$, and let $\alpha > 0$. Then,*

$$\Pr[|X - \mu| \geq \alpha] \leq \left(\frac{nk}{\alpha^2}\right)^{k/2}.$$

### 2.5.2 The Poisson Approximation

We will use the following useful facts about the Poisson approximation. When throwing $n$ balls into $R$ bins, the distribution of the number of balls in a given bin is $\mathrm{Bin}(n, 1/R)$. As the Poisson distribution is the limit distribution of the binomial distribution, the distribution of the number of balls in a given bin is approximately $\mathrm{Pois}(n/R)$. In fact, in some cases we could approximate the *joint distribution* of the number of balls in all the bins by assuming the load at each bin is an independent Poisson random variable with mean $n/R$.

**Theorem 2.4** (e.g., [18]). *Suppose that $n$ balls are thrown into $R$ bins independently and uniformly at random, and let $X_i$ be the number of balls in the $i^{th}$ bin, where $1 \leq i \leq R$. Let $Y_1, \cdots, Y_R$ be independent Poisson random variables with mean $n/R$. Let $f(x_1, \cdots, x_R)$ be a nonnegative function. Then,*

$$\mathbb{E}\left[f(X_1, \cdots, X_R)\right] \leq e\sqrt{n}\mathbb{E}\left[f(Y_1, \cdots, Y_R)\right].$$

In particular, the theorem states that any event that takes place with probability $p$ in the Poisson case, takes place with probability at most $pe\sqrt{n}$ in the exact case (this follows by letting $f$ be the indicator function of that event).

We will also use the following bounds for the tail probabilities of a Poisson random variable:

**Theorem 2.5** ([3]). *Let $X$ have Poisson distribution with mean $\mu$. For $0 \leq \alpha \leq 1$,*

$$\begin{aligned} \Pr[X \leq \mu(1-\alpha)] &\leq e^{-\alpha^2\mu/2} \\ \Pr[X \geq \mu(1+\alpha)] &\leq e^{-\alpha^2\mu/2}. \end{aligned}$$

# 3 Our Algorithms

In this section we give an informal description of our algorithms, and highlight some of the ideas behind our constructions.

## 3.1 The TreeHist Protocol

We briefly give an overview of our construction that is based on a compressed, noisy version of the count sketch. To maintain clarity of the main ideas, we give here a high-level description of our construction. We refer to Section 5 for a detailed description of this construction.

We first introduce some objects and public parameters that will be used in the construction:

**Prefixes:** For a binary string $v$, we will use $v[1 : \ell]$ to denote the $\ell$-bit prefix of $v$. Let $\overline{\mathcal{V}} = \{v \in \{0,1\}^\ell \text{ for some } \ell \in [\log d]\}$. Note that elements of $\overline{\mathcal{V}}$ arranged in a binary prefix tree of depth $\log d$, where the nodes at level $\ell$ of the tree represent all binary strings of length $\ell$. The items of the dictionary $\mathcal{V}$ represent the bottommost level of that tree. Note that $|\overline{\mathcal{V}}| = 2d$. We will use $\perp$ to denote an empty string. For a $v \in \{0,1\}^\ell$ and $b \in \{0,1\}$, let $v\|b$ denote the $\ell+1$-bit string resulting from appending the bit $b$ to $v$. For a binary string $v$, we define $\mathsf{Child}(v) \triangleq \{v\|0, v\|1\}$, that is, the set containing the two children of $v$ in the prefix tree. Similarly, for a set of strings $\mathcal{U}$, we define $\mathsf{ChildSet}(\mathcal{U}) \triangleq \{v : v \in \mathsf{Child}(u) \text{ for some } u \in \mathcal{U}\}$.

**Hashes:** Let $t, m$ be positive integers to be specified later. We will consider a set of $t$ pairs of hash functions $\{(h_1, g_1), \ldots, (h_t, g_t)\}$, where for each $i \in [t]$, $h_i : \overline{\mathcal{V}} \to [m]$ and $g_i : \overline{\mathcal{V}} \to \{-1, +1\}$ are independently and uniformly chosen pairwise independent hash functions.

**Basis matrix:** Let $\mathbf{W} \in \{-1, +1\}^{m \times m}$ be $\sqrt{m} \cdot \mathbf{H}_m$ where $\mathbf{H}_m$ is the Hadamard transform matrix of size $m$. As will be shown later, we will be making operations over the entries of this matrix. It is important to note that we do not need to store this matrix. The value of any entry in this matrix can be computed in $O(\log m)$ bit operations given the (row, column) index of that entry. In particular, suppose we want to compute the value of the entry $W_{i,j}$ located at the $i$-th row and $j$-th column. Let $(i_0, i_1, \ldots, i_{\log m - 1})$ and $(j_0, j_1, \ldots, j_{\log m - 1})$ denote the bit representation of $i$ and $j$, respectively. Then, $W_{i,j} = (-1)^{\sum_{k=0}^{\log m - 1} i_k j_k}$.

**Global parameters:** The total number of users $n$, the size of the Hadamard matrix $m$, the number of hash pairs $t$, the privacy parameter $\epsilon$, the confidence parameter $\beta$, and the hash functions $\{(h_1, g_1), \ldots, (h_t, g_t)\}$ are assumed to be public information. We set $t = O(\log(n/\beta))$ and $m = O\left(\sqrt{\frac{n}{\log(n/\beta)}}\right)$.

**Public randomness:** In addition to the $t$ hash pairs $\{(h_1, g_1), \ldots, (h_t, g_t)\}$, we assume that the server creates a random partition $\Pi : [n] \to [\log d] \times [t]$ that assigns to each user $i \in [n]$ a random pair $(\ell_i, j_i) \leftarrow [\log(d)] \times [t]$, and another random function $\mathcal{Q} : [n] \leftarrow [m]$ that assigns[3] to each user $i$ a uniformly random index $r_i \leftarrow [m]$. We assume that such random indices $\ell_i, j_i, r_i$ are shared between the server and each user.

First, we describe the two main modules of our protocol.

### 3.1.1 A local randomizer: LocalRnd

For each $i \in [n]$, user $i$ runs her own independent copy of a local randomizer, denoted as LocalRnd, to generate her private report. LocalRnd of user $i$ starts by acquiring the index triple $(\ell_i, j_i, r_i) \leftarrow [\log d] \times [t] \times [m]$ from public randomness. For each user, LocalRnd is invoked twice in the full protocol: once during the first phase of the protocol (called the pruning phase) where the high-frequency items (*heavy hitters*) are identified, and a second time during the final phase (the estimation phase) to enable the protocol to get better estimates for the frequencies of the heavy hitters. In the first invocation, LocalRnd of user $i$ performs its computation on the $\ell_i$-th prefix of the item $v_i$ of user $i$, whereas in the second invocation, it performs the computation on the entire user's string $v_i$.

Apart from this, in both invocations, LocalRnd follows similar steps. It first selects the hash pair $(h_{j_i}, g_{j_i})$, computes $c_i = h_{j_i}(v_i[1 : \tilde{\ell}])$ (where $\tilde{\ell} = \ell_i$ in the first invocation and $\tilde{\ell} = \log d$ in the second invocation, and $v_i[1 : \tilde{\ell}]$ is the $\tilde{\ell}$-th prefix of $v_i$), then it computes a bit $x_i = g_{j_i}\left(v_i[1:\tilde{\ell}]\right) \cdot W_{r_i, c_i}$ (where $W_{r,c}$ denotes the $(r, c)$ entry of the basis matrix $\mathbf{W}$). Finally, to guarantee $\epsilon$-local differential privacy, it generates a randomized response $y_i$ based on $x_i$ (i.e., $y_i = x_i$ with probability $e^{\epsilon/2}/(1 + e^{\epsilon/2})$ and $y_i = -x_i$ with probability $1/(1 + e^{\epsilon/2})$, which is sent to the server.

---
[3]We could have grouped $\Pi$ and $\mathcal{Q}$ into one random function mapping $[n]$ to $[\log d] \times [t] \times [m]$, however, we prefer to split them for clarity of exposition as each source of randomness will be used for a different role.

Our local randomizer can thought of as a transformed, compressed (via sampling), and randomized version of the count sketch [9]. In particular, we can think of LocalRnd as follows. It starts off with similar steps to the standard count sketch algorithm, but then deviates from it as it applies Hadamard transform to the user's signal, then samples one bit from the result. By doing so, we can achieve significant savings in space and communication without sacrificing accuracy.

### 3.1.2 A frequency oracle: FreqOracle

Suppose we want to allow the server estimate the frequencies of some *given* subset $\widehat{\mathcal{V}} \subseteq \{0, 1\}^\ell$ for some given $\ell \in [\log d]$ based on the noisy users' reports. We give a protocol, denoted as FreqOracle, for accomplishing this task.

For each queried item $\hat{v} \in \widehat{\mathcal{V}}$ and for each hash index $j \in [t]$, FreqOracle computes $c = h_j(\hat{v})$, then collects the noisy reports of a collection of users $\mathcal{I}_{\ell,j}$ that contains every user $i$ whose pair of prefix and hash indices $(\ell_i, j_i)$ match $(\ell, j)$. Next, it estimates the inverse Hadamard transform of the compressed and noisy signal of each user in $\mathcal{I}_{\ell,j}$. In particular, for each $i \in \mathcal{I}_{\ell,j}$, it computes $y_i \, W_{r_i, c}$ which can be described as a multiplication between $y_i \mathbf{e}_{r_i}$ (where $\mathbf{e}_{r_i}$ is the indicator vector with 1 at the $r_i$-th position) and the scaled Hadamard matrix $\mathbf{W}$, followed by selecting the $c$-th entry of the resulting vector. This brings us back to the standard count sketch representation. It then sums all the results and multiplies the outcome by $g_j(\hat{v})$ to obtain an estimate $\hat{f}_j(\hat{v})$ for the frequency of $\hat{v}$. As in the count sketch algorithm, this is done for every $j \in [t]$, then FreqOracle obtains a high-confidence estimate by computing the median of all the $t$ frequency estimates.

### 3.1.3 The protocol: TreeHist

The protocol is easier to describe via operations over nodes of the prefix tree $\overline{\mathcal{V}}$ of depth $\log d$ (described earlier). The protocol runs through two main phases: the pruning (or, scanning) phase, and the final estimation phase.

In the pruning phase, the protocol scans the levels of the prefix tree starting from the top level (that contains just 0 and 1) to the bottom level (that contains all items of the dictionary). For a given node at level $\ell \in [\log d]$, using FreqOracle as a subroutine, the protocol gets an estimate for the frequency of the corresponding $\ell$-bit prefix. For any $\ell \in [\log(d) - 1]$, before the protocol moves to level $\ell + 1$ of the tree, it prunes all the nodes in level $\ell$ that *cannot* be prefixes of actual heavy hitters (high-frequency items in the dictionary). Then, as it moves to level $\ell + 1$, the protocol considers only the children of the surviving nodes in level $\ell$. The construction guarantees that, with high probability, the number of surviving nodes in each level cannot exceed $O\left(\sqrt{\frac{n}{\log(d)\log(n)}}\right)$. Hence, the total number of nodes queried by the protocol (i.e., submitted to FreqOracle) is at most $O\left(\sqrt{\frac{n \log(d)}{\log(n)}}\right)$.

In the second and final phase, after reaching the final level of the tree, the protocol would have already identified a list of the candidate heavy hitters, however, their estimated frequencies may not be as accurate as we desire due to the large variance caused by the random partitioning of users across all the levels of the tree. Hence, it invokes the frequency oracle once more on those particular items, and this time, the sampling variance is reduced as the set of users is partitioned only across the $t$ hash pairs (rather than across $\log(d) \times t$ bins as in the pruning phase). By doing this, the server obtains more accurate estimates for the frequencies of the identified heavy hitters. The privacy and accuracy guarantees are stated below. The full details are given in Section 5.

### 3.1.4 Privacy and Utility Guartantees

**Theorem 3.1.** Protocol TreeHist is $\epsilon$-local differentially private.

**Theorem 3.2.** There is a number $\eta = O\left(\sqrt{n \log(n/\beta) \log(d)}/\epsilon\right)$ such that with probability at least $1 - \beta$, the output list of the TreeHist protocol satisfies the following properties:

1. it contains all items $v \in \mathcal{V}$ whose true frequencies above $3\eta$.

2. it does not contain any item $v \in \mathcal{V}$ whose true frequency below $\eta$.

3. Every frequency estimate in the output list is accurate up to an error $\leq O\left(\sqrt{n \log(n/\beta)}/\epsilon\right)$

## 3.2 The `Bitstogram` Protocol

We now present a simplified description of our second protocol, that captures most of the ideas. Any informalities made hereafter are removed in the full description of the protocol (Section 6).

**First Step: Frequency Oracle.** Recall that a frequency oracle is a protocol that, after communicating with the users, outputs a *data structure* capable of approximating the frequency of every domain element $v \in \mathcal{V}$. So, if we were to allow the server to have linear runtime in the domain size $|\mathcal{V}| = d$, then a frequency oracle would suffice for computing histograms. As we are interested in protocols with a significantly lower runtime, we will only use a frequency oracle as a subroutine, and query it only for (roughly) $\sqrt{n}$ elements.

Let $Z \in \{\pm 1\}^{d \times n}$ be a matrix chosen uniformly at random, and assume that $Z$ is publicly known.[4] That is, for every domain element $v \in \mathcal{V}$ and every user $j \in [n]$, we have a random bit $Z[v, j] \in \{\pm 1\}$. As $Z$ is publicly known, every user $j$ can identify its corresponding bit $Z[v_j, j]$, where $v_j \in \mathcal{V}$ is the input of user $j$. Now consider a protocol in which users send randomized responses of their corresponding bits. That is, user $j$ sends $y_j = Z[v_j, j]$ w.p. $\frac{1}{2} + \frac{\epsilon}{2}$ and sends $y_j = -Z[v_j, j]$ w.p. $\frac{1}{2} - \frac{\epsilon}{2}$. We can now estimate the frequency of every domain element $v \in \mathcal{V}$ as

$$a(v) = \frac{1}{\epsilon} \cdot \sum_{j \in [n]} y_j \cdot Z[v, j].$$

To see that $a(v)$ is accurate, observe that $a(v)$ is the sum of $n$ independent random variables (one for every user). For the users $j$ holding the input $v$ (that is, $v_j = v$) we will have that $\frac{1}{\epsilon}\mathbb{E}[y_j \cdot Z[v, j]] = 1$. For the other users we will have that $y_j$ and $Z[v, j]$ are independent, and hence $\mathbb{E}[y_j \cdot Z[v, j]] = \mathbb{E}[y_j] \cdot \mathbb{E}[Z[v, j]] = 0$. That is, $a(v)$ can be expressed as the sum of $n$ independent random variables: $f(v)$ variables with expectation 1, and $(n - f(v))$ variables with expectation 0. The fact that $a(v)$ is an accurate estimation for $f(v)$ now follows from the Hoeffding bound.

**Lemma 3.3** (Algorithm `Hashtogram`). *Let $\epsilon \leq 1$. Algorithm `Hashtogram` satisfies $\epsilon$-LDP. Furthermore, with probability at least $1 - \beta$, algorithm `Hashtogram` answers every query $v \in \mathcal{V}$ with $a(v)$ satisfying: $|a(v) - f(v)| \leq O\left(\frac{1}{\epsilon} \cdot \sqrt{n \log\left(\frac{nd}{\beta}\right)}\right)$.*

**Second Step: Identifying Heavy-Hitters.** Let us assume that we have a frequency oracle protocol with worst-case error $\tau$. We now want to use our frequency oracle in order to construct a protocol that operates on two steps: First, it identifies a small set of potential "heavy-hitters", i.e., domain elements that appear in the database at least $2\tau$ times. Afterwards, it uses the frequency oracle to estimate the frequencies of those potential heavy elements.[5]

Let $h : \mathcal{V} \to [T]$ be a (publicly known) random hash function, mapping domain elements into $[T]$, where $T$ will be set later.[6] We will now use $h$ in order to identify the heavy-hitters. To that end, let $v^* \in \mathcal{V}$ denote such a heavy-hitter, appearing at least $2\tau$ times in the database $S$, and denote $t^* = h(v^*)$. Assuming that $T$ is big enough, w.h.p. we will have that $v^*$ is the only input element (from $S$) that is mapped (by $h$) into the hash value $t^*$. Assuming that this is indeed the case, we will now identify $v^*$ bit by bit.

For $\ell \in [\log d]$, denote $S_\ell = (h(v_j), v_{j,\ell})_{j \in [n]}$, where $v_{j,\ell}$ is bit $\ell$ of $v_j$. That is, $S_\ell$ is a database over the domain $([T] \times \{0, 1\})$, where the row corresponding to user $j$ is $(h(v_j), v_{j,\ell})$. Observe that every user can compute her own row locally. As $v^*$ is a heavy-hitter, for every $\ell \in [\log d]$ we have that $(t^*, v_\ell^*)$ appears in $S_\ell$ at least $2\tau$ times, where $v_\ell^*$ is bit $\ell$ of $v^*$. On the other hand, as we assumed that $v^*$ is the only input element that is mapped into $t^*$ we get that $(t^*, 1 - v_\ell^*)$ does not appear in $S_\ell$ at all. Recall that our frequency oracle has error at most $\tau$, and hence, we can use it to accurately determine the bits of $v^*$.

---

[4] As we later explain, $Z$ has a short description, as it need not be uniform.

[5] Event though we describe the protocol as having two steps, the necessary communication for these steps can be done in parallel, and hence, our protocol will have only 1 round of communication.

[6] As with the matrix $Z$, the hash function $h$ can have a short description length.

To make things more concrete, consider the protocol that for every hash value $t \in [T]$, for every coordinate $\ell \in [\log d]$, and for every bit $b \in \{0, 1\}$, obtains an estimation (using the frequency oracle) for the duplicity of $(t, b)$ in $S_\ell$ (so there are $\log d$ invocations of the frequency oracle, and a total of $2T \log d$ estimations). Now, for every $t \in [T]$ let us define $\hat{v}_t$ where bit $\ell$ of $\hat{v}_t$ is the bit $b$ s.t. $(t, b)$ is more frequent than $(t, 1 - b)$ in $S_\ell$. By the above discussion, we will have that $\hat{v}_{t^*} = v^*$. That is, the protocol identifies a set of $T \log d$ domain elements, containing all of the heavy-hitters. The frequency of the identified heavy-hitters can then be estimated using the frequency oracle.

**Remark 3.1.** As should be clear from the above discussion, it suffices to take $T \gtrsim n^2$, as this will ensure that there are no collisions among different input elements. As we only care about collisions between "heavy-hitters" (appearing in $S$ at least $\sqrt{n}$ times), it would suffice to take $T \gtrsim n$ to ensure that w.h.p. there are no collisions between heavy-hitters. In fact, we could even take $T \gtrsim \sqrt{n}$, which would ensure that a heavy-hitter $v^*$ has no collisions with constant probability, and then to amplify our confidence using repetitions.

**Lemma 3.4** (Algorithm `Bitstogram`)**.** *Let $\epsilon \leq 1$. Algorithm `Bitstogram` satisfies $\epsilon$-LDP. Furthermore, the algorithm returns a list $L$ of length $\tilde{O}(\sqrt{n})$ satisfying:*

1. *With probability $1 - \beta$, for every $(v, a) \in L$ we have that $|a - f(v)| \leq O\left(\frac{1}{\epsilon}\sqrt{n \log(n/\beta)}\right)$.*
2. *W.p. $1 - \beta$, for every $v \in \mathcal{V}$ s.t. $f(v) \geq O\left(\frac{1}{\epsilon}\sqrt{n \log(d/\beta) \log(\frac{1}{\beta})}\right)$, we have that $v$ is in $L$.*

## 4 Detailed Experimental Results

In this section we discuss implementation details of our algorithms mentioned in Section 5[7]. The main objective of this section is to emphasize the empirical efficacy of our algorithms along with the theoretical optimality in terms of error, space, time and communication. [19] recently claimed space space optimality for a similar problem, but a formal analysis (or empirical evidence) was not provided. Our experiments corroborate both the analytical bounds in our current work, and in [19]. Our experiments are performed on a macOS-Sierra 10.12 system (in Python 2.7) with 3.3Ghz (Intel Core i5) and 16GB of DDR-3 RAM.

### 4.1 Private Frequency Oracle

In this experiment, the objective is to test the efficacy of our algorithm in estimating the frequencies of a known set of dictionary of user items, under local differential privacy. We estimate the error in estimation while varying the size of the data set $n$, changing the privacy parameter $\epsilon$. (See Section 2.1 for a refresher on the notation.)

Figure 1 shows results on a synthetic data set with the domain size of hundred (i.e., $d = 100$) drawn from a power law distribution with power of $15$. The default parameters used in Figure 1 are: *number of data samples* $(n)$ : 10 million, range of the hash function $(m)$: $\sqrt{n}$, number of hash functions $(t)$: 285, and the privacy parameter $\epsilon = 2.0$. For the hash functions, we used the prefix bits of SHA-256. The estimated frequency is scaled by the number of samples to normalize the result, and each experiment is averaged over *ten runs*. The bars for True refers to the the true frequencies, and the bars for Priv corresponds to the differentially private frequencies. The pctle corresponds to the frequency of a domain element at the corresponding percentile in the frequency distribution of the data set. *Observations:* i) The plots corroborate the fact that the frequency oracle is indeed *unbiased*. The average frequency estimate (over ten runs) for each percentile is within one standard deviation of the corresponding true estimate. ii) The error in the estimates go down significantly as the number of samples are increased or the privacy parameter $\epsilon$ is increased.

In Figure 2 we show the result of changing the range of the hash function ($m$). The observation is that the results are seemingly *insensitive* to the range of the hash function.

---

[7]The experiments are performed without the Hadamard compression during data transmission.

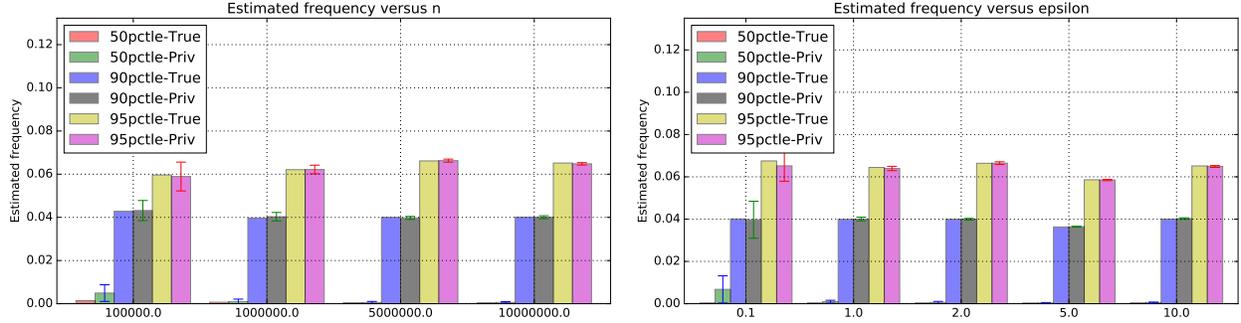

Figure 1: Frequency vs number of samples ($n$) and privacy parameter ($\epsilon$) on the synthetic data set.

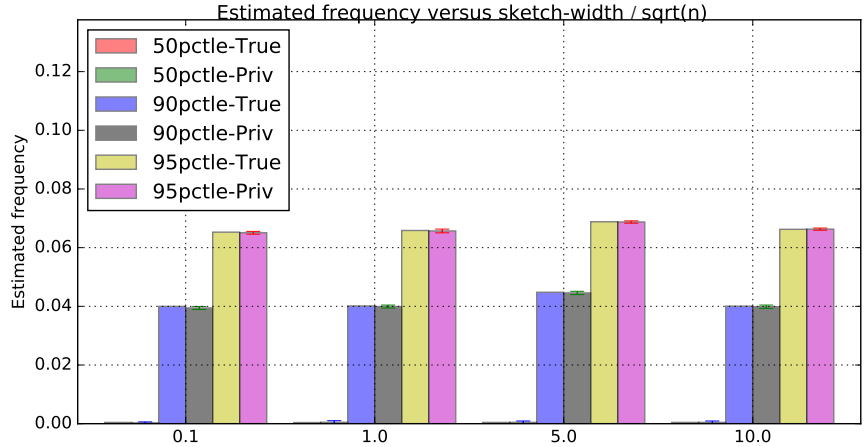

Figure 2: Frequency vs sketch width ($m$) on the synthetic data set.

We also ran the same experiment (Figure 3) on a real data set drawn uniformly at random the NLTK Brown corpus [1]. The data set we created has $n = 10$ million samples drawn i.i.d. from the corpus with replacement, and the system parameters are the same default parameters described earlier. In this plot, the rank corresponds to the rank of a domain element in the distribution of frequencies in the data set. The observation is here is also consistent with that of Figure 1.

***Comparison to RAPPOR [11]:*** Here we compare ourselves to the only other system (RAPPOR project from GOOGLE) for the private frequency estimation problem whose code is publicly available. We took the snapshot of their code base (https://github.com/google/rappor) on May 9th, 2017. In order to perform a fair comparison, we tested our algorithm against one of their demo experiments available (*Demo3* using the demo.sh script). We used the privacy parameter $\epsilon = \ln(3)$, the number of data samples $n = 1$ million, and the data set to be the same data set generated by the demo.sh script. In Figure 4 we observe that for higher frequencies both RAPPOR and our algorithm perform similarly. However, in lower frequency regimes, the RAPPOR estimates are zero most of the times, while our estimates are closer to the true estimates. *N.B.* We do not claim that our algorithm would outperform the RAPPOR system on all problem instances. However, our current experiment does highlight the need to perform an *at-scale* comparison between the two algorithms.

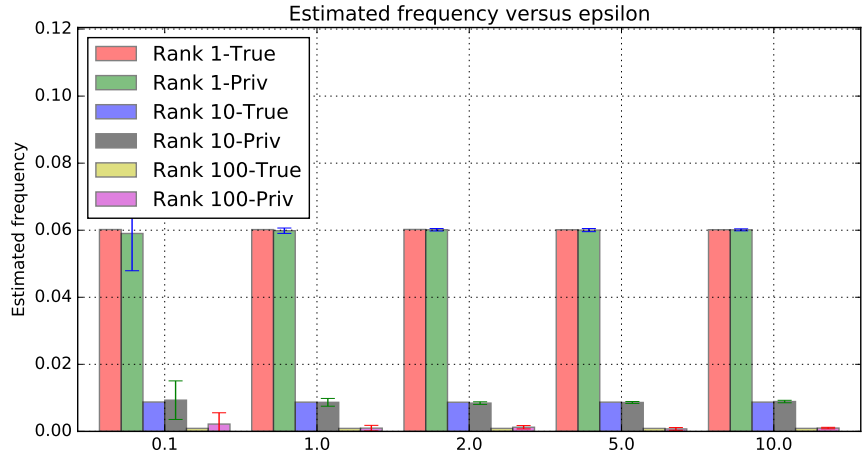

Figure 3: Frequency vs privacy ($\epsilon$) on the NLTK-Brown corpus.

## 4.2 Private Heavy-hitters

In this section, we take on the harder task of identifying the heavy hitters, rather than estimating the frequencies of domain elements. We run our experiments on the same two data sets described earlier with the same setting of parameters, except now we assume that we do not know the domain. As a part of our algorithm design, we assume that every element in the domain is from the english alphabet set [a-z] and are of length exactly equal to six. If they are of larger length, we truncate then before entering them in the data set, and if they are of smaller length we tag a $\perp$ at the end. We generate the domain elements for the synthetic data set by first generating the frequency histogram based on the power law distribution described earlier, and then assign random strings of length eight to each bin of the histogram. That becomes our data set. For the NLTK Brown corpus [1], we sample $n = 10$ million samples with replacement from the corpus to form our data set. We set a threhold of $15 \cdot \sqrt{n}$ as the threshold for being a heavy hitter. We measure the efficacy of our system by measuring the *precision* and *recall*. Figures 5 and 6 show the true data distribution for the synthetic and the NLTK data set.

In Table 4.2 we state our corresponding precision and recall parameters. Our recall numbers are much better than the precision numbers, primarily because of the large number of negative examples ($3 \times 10^8$ examples). In practice, if there are *false-positives*, they can be easily pruned using domain expertise. For example, if we are trying to identify new words which users are typing in English [2], then using the domain expertise of English, a set of false positives can be easily ruled out by inspecting the list of heavy hitters output by the algorithm. Further, notice that since we are working with domain elements with size six characters, a brute force algorithm would require $26^6$ queries to the frequency oracle, which would be computationally (near) infeasible. While there are other algorithms for finding heavy-hitters [4, 15], either they do not provide any theoretical guarantee for the utility [11, 13, 19], or there does not exist a scalable and efficient implementation for them. Our work scores well on both these aspects.

| Data set | # of unique words | Precision | Recall |
|---|---|---|---|
| Synthetic | 93 | 0.36 ($\sigma = 0.05$) | 0.95 ($\sigma = 0.03$) |
| NLTK Brown corpus | 25991 | 0.24 ($\sigma = 0.04$) | 0.86 ($\sigma = 0.05$) |

Table 2: Private Heavy-hitters with threshold=$15\sqrt{n}$. Here $\sigma$ corresponds to the standard deviation.

## 5 Locally Private Heavy-hitters via Count-Sketch: The TreeHist Protocol

We start with a detailed description of our construction described at a high level in Section 3.1.

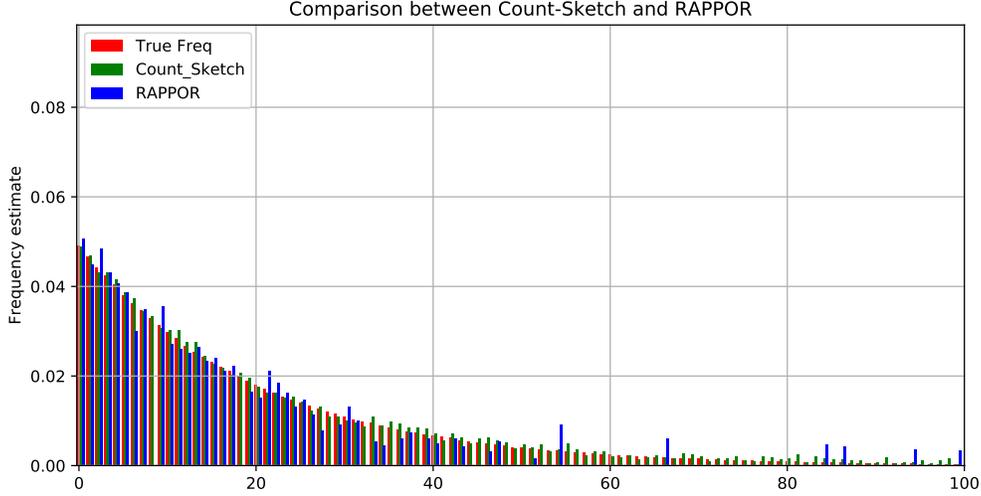

Figure 4: Frequency vs privacy ($\epsilon$) on the Demo 3 experiment from RAPPOR

We refer to Section 3.1 for definitions and public parameters that will be used in the construction, namely, *prefixes, hashes, the basis matrix, global parameters, and public randomness*. We restate below our public parameters and, when applicable, their specific settings.

**Global parameters:** We will assume that total number of users $n$, the size of the Hadamard matrix $m$, the number of hash pairs $t$, the privacy parameter $\epsilon$, and the confidence parameter $\beta$ are public parameters, and hence they will not be explicitly provided as inputs to the algorithms. For integer parameters, we will implicitly assume that results are rounded to the nearest integer. We set $t = 110 \log(n/\beta)$ and $m = 48\sqrt{\frac{n}{\log(n/\beta)}}$. The hash functions $\{(h_1, g_1), \ldots, (h_t, g_t)\}$ will also be assumed to be public information (this is $O(\log(d) \log(n/\beta))$ bits of shared randomness).

**Public randomness:** In addition to the $t$ hash pairs $\{(h_1, g_1), \ldots, (h_t, g_t)\}$, we assume that the server creates a random partition $\Pi : [n] \to [\log d] \times [t]$ over the set of users, that is, a each user $i$ gets a random pair $(\ell_i, j_i) \leftarrow [\log(d)] \times [t]$ that represents the index of one of $\log(d) \times t$ "buckets." Moreover, the server uses another random function $\mathcal{Q} : [n] \leftarrow [m]$ that assigns to each user $i$ a uniformly random index $r_i \leftarrow [m]$. We assume that such random indices $\ell_i, j_i, r_i$ are shared between the server and each user. For each $\ell \in [\log d]$ and each $j \in [t]$, we define $\mathcal{I}_{\ell,j} \triangleq \{i : \ell_i = \ell, j_i = j\}$ and $\mathcal{I}_j \triangleq \{i : j_i = j\} = \cup_{\ell \in [\log d]} \mathcal{I}_{\ell,j}$.

## 5.1 A Local Randomizer: LocalRnd

For each $i \in [n]$, user $i$ runs her own independent copy of Algorithm 1 below, refered to as LocalRnd, to generate her private report. We note that LocalRnd takes a flag Final $\in \{0, 1\}$ as an input. The role of this input will become clear when we discuss the full protocol. In a nutshell, the flag is used to distinguish between two invocations of LocalRnd. In particular, the local randomizer of each user is invoked twice in the full protocol: once during the first phase of the protocol (called the pruning phase) where the high-frequency items (*heavy hitters*) are identified, and a second time during the final phase (the final estimation phase) to enable the protocol to get better estimates for the frequencies of the heavy hitters.

**Connection to count sketch and Hadamard transform:** Our local randomizer can thought of as a transformed, compressed (via sampling), and randomized version of the count sketch. Up to Step 5 in LocalRnd (Algorithm 1), our algorithm follows the standard count sketch algorithm [9]. Starting from Step 6, we start to deviate from the standard count sketch as we apply Hadamard transform to the user's signal, then sample one bit from the result. Indeed Step 6 can be thought of as a composition of two operations: first, we multiply the indicator vector $\mathbf{e}_{c_i} \in \{0, 1\}^m$ by the scaled Hadamard matrix $\mathbf{W}$ (this is equivalent to selecting $\mathbf{w}_{c_i}$: the $c_i$-th column of $\mathbf{W}$), then we randomly sample one entry from $\mathbf{w}_{c_i}$. By doing so, we can achieve significant savings in space and communication without sacrificing accuracy.

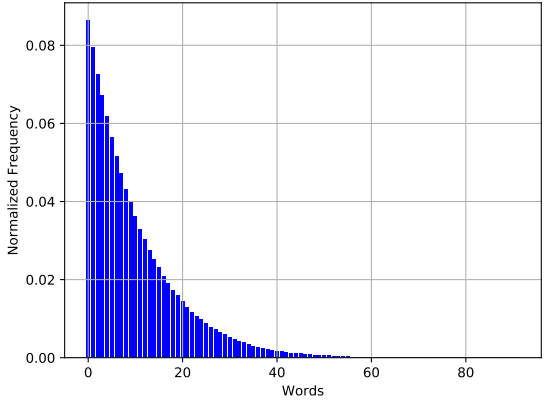

Figure 5: Distr. for Synthetic

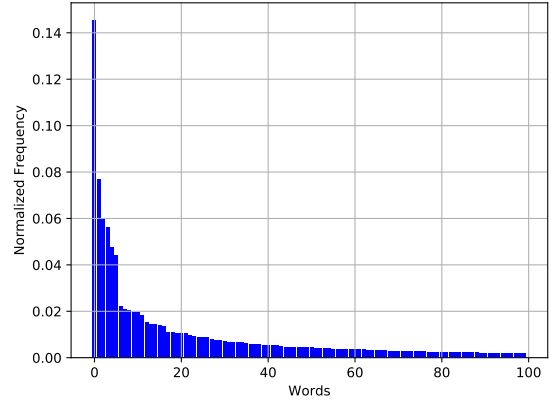

Figure 6: Distr. for Brown (top 100 words)

---

**Algorithm 1** LocalRnd

---

**Input:** User $i$ input: $v_i \in \mathcal{V}$, Flag: Final $\in \{0, 1\}$.
1: Using shared randomness, get random indices $(\ell_i, j_i) \leftarrow [\log d] \times [t]$ and $r_i \leftarrow [m]$.
2: **if** Final $= 0$ **then**
3:     Set $s_i := g_{j_i}(v_i[1 : \ell_i])$ and $c_i := h_{j_i}(v_i[1 : \ell_i])$. $\{v[1 : \ell]$ denotes the $\ell$-bit prefix of $v$.$\}$
4: **else**
5:     Set $s_i := g_{j_i}(v_i)$ and $c_i := h_{j_i}(v_i)$
6: Compute $x_i := s_i \cdot W_{r_i, c_i}$
    $\{W_{r, c}$ denotes the sign of the $(r, c)$ entry of $\mathbf{H}_m$ (Hadamard matrix of size $m$).$\}$
7: Generate a randomized version $y_i$ of the bit $x_i$:

$$y_i = \begin{cases} x_i & \text{w.p. } \frac{e^{\epsilon/2}}{e^{\epsilon/2}+1} \\ -x_i & \text{w.p. } \frac{1}{e^{\epsilon/2}+1} \end{cases}$$

8: **return** $y_i$.

---

The output of LocalRnd is 1 bit per invocation. Hence, during the entire span of the protocol, each user sends 2 bits to the server. Here, we will assume the user's identity (i.e., the index $i \in [n]$) associated with each report is known to the server, (e.g., from a higher layer in the communication protocol stack).

## 5.2 A Frequency Oracle: FreqOracle

Before describing our protocol for identifying the heavy hitters and estimating their frequencies, we first discuss a protocol for a simpler task. Suppose we want to allow the server estimate the frequencies of some *given* subset $\widehat{\mathcal{V}} \subseteq \{0, 1\}^\ell$ for some given $\ell \in [\log d]$ based on the users' reports. Algorithm 2 describes a protocol, denoted as FreqOracle, for accomplishing this task. Note that in this protocol, we assume that all items whose frequencies are in question are given as inputs.

For each queried item $\hat{v} \in \widehat{\mathcal{V}}$ and for each hash index $j \in [t]$, FreqOracle starts by collecting the noisy reports of the collection $\mathcal{I}_{\ell, j}$ of users where each user $i \in \mathcal{I}_{\ell, j}$ is assigned a pair of prefix and hash indices $(\ell_i, j_i)$ that matches $(\ell, j)$. Next, it estimates the inverse Hadamard transform of the compressed and noisy signal of each user in $\mathcal{I}_{\ell,j}$. It then sums all the results and multiplies the outcome by $g_j(\hat{v})$ to obtain an estimate $\hat{f}_j(\hat{v})$ for the frequency of $\hat{v}$. As in the count sketch algorithm, this is done for every $j \in [t]$, then FreqOracle obtains a high-confidence estimate for the frequency of $\hat{v}$ by computing the median of all the $t$ frequency estimates.

**Inverse transform and back to count sketch:** We note here that Steps 7 in FreqOracle (Algorithm 2) can be described as an inverse Hadamard transform of the users' compressed and noisy signals. In particular, each term $y_i\, W_{r_i, c}$ inside the sum can be described as a multiplication between $y_i \mathbf{e}_{r_i}$ (where $\mathbf{e}_{r_i}$ is the indicator vector with 1 at the $r_i$-th position) and the scaled Hadamard matrix $\mathbf{W}$, followed by picking the $c$-th entry of the resulting vector. This brings us back to the standard count sketch representation [9]. The FreqOracle protocol then proceeds to process the frequency estimates in the same way a count sketch does. Indeed, Step 8 is a standard step in count sketch algorithm where a high-confidence frequency estimate is obtained via the median technique. Hence, we attain the functionality of the count sketch algorithm with much less space and communication by transforming the users' signals to the Fourier domain, compressing signals via sampling, and then transforming them back at the server.

---

**Algorithm 2** FreqOracle

---

**Input:** Prefix length: $\ell \in [\log d]$, a subset of $\ell$-bit prefixes $\widehat{\mathcal{V}} \subseteq \{0,1\}^\ell$, collection of $t$ disjoint subsets of users: $\left\{ \tilde{\mathcal{I}}_j\, :\, j \in [t] \right\}$, oracle access to items of relevant users: $\left( v_i\, :\, i \in \cup_{j \in [t]} \tilde{\mathcal{I}}_j \right)$, scaling factor: $\gamma$, Flag: Final $\in \{0, 1\}$. $\{ a_\epsilon \triangleq \frac{e^{\epsilon/2}+1}{e^{\epsilon/2}-1}$ is assumed to be a global constant seen by FreqOracle.$\}$

1: **for** $\hat{v} \in \widehat{\mathcal{V}}$ **do**
2:    **for** Hash index $j = 1$ to $t$ **do**
3:       Set $s := g_j(\hat{v})$ and $c := h_j(\hat{v})$.
4:       **for** Users $i \in \tilde{\mathcal{I}}_j$ **do**
5:          Get user-$i$'s 1-bit report: $y_i = \mathsf{LocalRnd}(v_i, \mathsf{Final})$
6:          Get user-$i$'s random index $r_i = \mathcal{Q}(i)$ using public randomness.
7:       Compute the $j$-th estimate of the frequency of $\hat{v}$: $\hat{f}_j(\hat{v}) := \gamma \cdot a_\epsilon \sum_{i \in \tilde{\mathcal{I}}_j} y_i \cdot s \cdot W_{r_i, c}$.
8:    Compute final estimate for the frequency of $\hat{v}$: $\hat{f}(\hat{v}) := \mathsf{Median}\left( \left\{ \hat{f}_j(\hat{v})\, :\, j \in [t] \right\} \right)$.
9: FreqList $:= \left\{ \left( \hat{v}, \hat{f}(\hat{v}) \right)\, :\, \hat{v} \in \widehat{\mathcal{V}} \right\}$.
10: **return** FreqList.

---

### 5.3 Succinct Histogram via a Tree Aggregation Protocol: TreeHist

We now describe our protocol (Algorithm 3), denoted as TreeHist, that outputs a succinct histogram, that is, it outputs a list of *heavy hitters* together with estimates for their frequencies. W.l.o.g., we will assume that $n > \log(n/\beta) \log(d)$ since otherwise, we cannot guarantee less than trivial error (i.e., an error of order $n$).

**The TreeHist protocol:**

    The protocol is easier to describe via operations over nodes of the prefix tree $\overline{\mathcal{V}}$ of depth $\log d$. The protocol runs through two main phases: the pruning (or, scanning) phase, and the final estimation phase.

    In the pruning phase, the protocol scans the levels of the prefix tree starting from the top level (that contains just 0 and 1) to the bottom level (that contains all items of the dictionary). For a given node at level $\ell \in [\log d]$, using FreqOracle as a subroutine, the protocol gets an estimate for the frequency of the corresponding $\ell$-bit prefix. As explained above, this estimate is obtained by FreqOracle by computing an estimate of the inverse Hadamard transform for the signal from each user that gets assigned to this level of the tree, and aggregating the resulting signals in exactly the same manner as it would be done in a standard count sketch (See Steps 7 and 8 in Algorithm 2). For any $\ell \in [\log(d) - 1]$, before the protocol moves to level $\ell + 1$ of the tree, it prunes all the nodes in level $\ell$ that *cannot* be prefixes of actual heavy hitters (high-frequency items in the dictionary). Then, as it moves to level $\ell + 1$, the protocol considers only the children of the surviving nodes in level $\ell$. The construction guarantees that, with high probability,

**Algorithm 3** TreeHist The Full Protocol
___
**Input:** Oracle access to users' items $(v_i \in \mathcal{V} : i \in [n])$.
1: Set $\eta := 147\sqrt{n \log(n/\beta) \log(d)}/\epsilon$ and $a_\epsilon := \frac{e^{\epsilon/2}+1}{e^{\epsilon/2}-1}$.
2: Set $\gamma := t \log d$ {This setting will change at the final stage of the protocol.}
3: Initialize Prefixes $= \{\bot\}$.
4: **for** Prefix length $\ell = 1$ to $\log d$ **do**
5:     Get the collection $\{\mathcal{I}_{\ell,j} : j \in [t]\}$ using the random partition $\Pi$. {See "Public randomness" above.}
6:     $\left\{\left(\hat{v}, \hat{f}(\hat{v})\right) : \hat{v} \in \mathsf{ChildSet}\,(\mathsf{Prefixes})\right\} =$

        $\mathsf{FreqOracle}\bigg(\ell, \mathsf{ChildSet}\,(\mathsf{Prefixes}), \{\mathcal{I}_{\ell,j} : j \in [t]\}, \left(v_i : i \in \cup_{j \in [t]}\mathcal{I}_{\ell,j}\right), \gamma, \mathsf{Final} = 0\bigg).$
7:     Initialize NewPrefixes $= \emptyset$.
8:     **for** $v \in \mathsf{ChildSet}\,(\mathsf{Prefixes})$ **do**
9:        **if** $\hat{f}(\hat{v}) \geq 2\eta$ **then**
10:           Add $\hat{v}$ to NewPrefixes.
11:     Update Prefixes $\leftarrow$ NewPrefixes.
12: Set $\gamma := t$
13: $\mathsf{SuccHist} := \mathsf{FreqOracle}\bigg(\log d, \mathsf{Prefixes}, \{\mathcal{I}_j : j \in [t]\}, \left(v_i : i \in [n]\right), \gamma, \mathsf{Final} = 1\bigg).$

    {Here $\mathcal{I}_j = \cup_{\ell \in [\log d]}\mathcal{I}_{\ell,j}$ as defined earlier in the "Public randomness" paragraph.}
14: **return** SuccHist.
___

the number of surviving nodes in each level cannot exceed $O\left(\sqrt{\frac{n}{\log(d)\log(n/\beta)}}\right)$. Hence, the total number of nodes queried by the protocol (i.e., submitted to FreqOracle) is at most $O\left(\sqrt{\frac{n\log(d)}{\log(n/\beta)}}\right)$.

In the second and final phase, after reaching the final level of the tree, the protocol would have already identified a list of potential heavy hitters, however, their estimated frequencies may not be as accurate as we desire due to the large variance caused by the random partitioning of users across all the levels of the tree. Hence, the protocol invokes the frequency oracle once more on those particular items, and this time, the sampling variance is reduced as the set of users is partitioned only across the $t$ hash pairs (rather than across $\log(d) \times t$ bins as in the pruning phase). By doing this, the server obtains more accurate estimates for the frequencies of the identified heavy hitters. The privacy and accuracy guarantees are formally stated in the following section.

## 5.4 Running Time and Processing Memory

**Running Time:** We note that Algorithm FreqOracle (Algorithm 2) is invoked $\log(d)+1 = O(\log d)$ times by TreeHist (Algorithm 3): once per each level of the tree and another at the final estimation phase at the bottom level of the tree. The main time-consuming step in FreqOracle is the summation in Step 7. This step is executed $O\left(\sqrt{\frac{n \log n}{\log d}}\right)$ times in every invocation of FreqOracle since there are $O\left(\sqrt{\frac{n}{\log n \log d}}\right)$ nodes queried in every level of the tree and for each node FreqOracle computes $t \approx \log n$ frequency estimates. That is, in total, this step is executed $O\left(\sqrt{n \log n \log d}\right)$ times in the TreeHist protocol. A direct implementation would involve summing $\approx n$ bits each time this step is executed, and hence would amount for running time of $\approx n^{1.5}$. However, we now show that this can be reduced to $\approx n$.

Consider Step 7 of Algorithm FreqOracle when FreqOracle is invoked by TreeHist at any given prefix level $\ell \in [\log d]$ during the pruning phase. We will not consider the final estimation phase here since its running time is dominated by that of the pruning ohase. Note that since the size of the basis matrix $\mathbf{W}$ is $m = O\left(\sqrt{\frac{n}{\log n}}\right)$, there are only $O\left(\sqrt{\frac{n}{\log n}}\right)$ values that the row index $r_i$ can take. Hence, $\hat{f}_j(\hat{v})$ (computed in Step 7) can be expressed as follows:

$$\hat{f}_j(\hat{v}) = \gamma \cdot a_\epsilon \sum_{\kappa \in [m]} \left( \sum_{i \in \mathcal{I}_{\ell,j}:\, r_i = \kappa} y_i \right) \cdot s \cdot W_{\kappa,c}.$$

Thus, to implement Step 7 of FreqOracle for all items $(\ell, j, \hat{v})$, we first compute $\left(\sum_{i \in \mathcal{I}_{\ell,j}:\, r_i = \kappa} y_i\right)$ for every $\kappa \in [m]$, $\ell \in [\log d]$, and $j \in [t]$ (this amounts to a running time of $O(n)$ in total). Then, for every value of $(\ell, j, \hat{v})$, computing $\hat{f}_j(\hat{v})$ would require summing $m = O\left(\sqrt{\frac{n}{\log n}}\right)$ numbers. Hence, in total, the running time of TreeHist is $O\left(\sqrt{\frac{n}{\log n \log d}} \cdot m \cdot t \log d\right) = O\left(n\sqrt{\log d}\right)$.

**Processing memory:** For the implementation described above, Algorithm FreqOracle maintains $m \cdot t \cdot \log d$ sums of at most $n$ bits each. This would require a processing memory of $O\left(\sqrt{n} \log^{1.5}(n) \log(d)\right)$ bits. The memory required for all the remianing steps of the TreeHist protocol does not exceed this amount, and hence, the total processing memory required by TreeHist is $O\left(\sqrt{n} \log^{1.5}(n) \log(d)\right)$ bits.

## 5.5 Privacy and Utility Guartantees

In this section we provide the privacy and utility gurantees for the TreeHist protocol.

**Theorem 5.1.** [$\epsilon$-Local Differential Privacy of the TreeHist Protocol] The TreeHist protocol (Algorithm 3) is $\epsilon$-local differentially private.

**Theorem 5.2.** [Utility of the TreeHist Protocol] Let $\eta = 147\sqrt{n \log(n/\beta) \log(d)}/\epsilon$. With probability at least $1 - \beta$, the output SuccHist of the TreeHist protocol (Algorithm 3) satisfies the following properties:

1. SuccHist contains all items $v \in \mathcal{V}$ whose true frequencies above $3\eta$.

2. SuccHist does not contain any item $v \in \mathcal{V}$ whose true frequency below $\eta$.

3. Every frequency estimate in SuccHist is accurate up to an error $\leq 147\frac{\sqrt{n \log(n/\beta)}}{\epsilon}$

We defer the proofs of Theorems 5.1 and 5.2 to Appendix A.

The following lemma state the error guarantees of the Algorithm FreqOracle used by the TreeHist protocol. Note that FreqOracle is invoked by the TreeHist protocol during both pruning and final estimation phases of the protocol. This lemma is central to our proof of Theorem 5.2.

**Lemma 5.3.** Let $\beta \in (0, 1)$. Let the number of hash pairs $t \geq 110 \log(n/\beta)$, and the size of Hadamard basis matrix $m \geq 48\sqrt{\frac{n}{\log(n/\beta)}}$. Consider algorithm FreqOracle (Algorithm 2) as invoked by the TreeHist protocol (Algorithm 3). For any $\ell \in [\log d]$, suppose that FreqOracle is invoked on inputs: $\ell$, subset $\widehat{\mathcal{V}} \subseteq \{0,1\}^\ell$ of size $|\widehat{\mathcal{V}}| \leq \sqrt{n}$, collection of users' subsets $\{\tilde{\mathcal{I}}_j : j \in [t]\}$, and scaling factor[8] $\gamma$. Assuming that $n \geq 48t \log(d)$, then, with probability $1 - \frac{\beta}{\log d}$, we have:

$$\forall v \in \widehat{\mathcal{V}}:\ |\hat{f}(v) - f(v)| 14\sqrt{n\gamma}/\epsilon.$$

where $\hat{f}(v)$ is the estimate of FreqOracle for the true frequency $f(v)$ of the item $v \in \widehat{\mathcal{V}}$.

*Proof.* Fix $\beta \in (0,1)$, $\ell \in [\log d]$, $t \geq 110 \log(n/\beta)$, $m \geq 48\sqrt{\frac{n}{\log(n/\beta)}}$, and a subset $\widehat{\mathcal{V}} \subseteq \{0,1\}^\ell$ of size $|\widehat{\mathcal{V}}| \leq \sqrt{n}$.

---

[8]Note that there are two possible settings for each of the collection $\{\tilde{\mathcal{I}}_j\}$ and the scaling factor $\gamma$: one for each invocation of FreqOracle by TreeHist, i.e., one for each value of the input flag Final. The generic notation $\{\tilde{\mathcal{I}}_j\}$ and $\gamma$ is used here to denote either of these two cases.

We note that when FreqOracle is invoked by TreeHist in the pruning phase (i.e., Final = 0), the collection $\{\tilde{\mathcal{I}}_j : j \in [t]\} = \{\mathcal{I}_{\ell,j} : j \in [t]\}$ and $\gamma = t \log d$, whereas when FreqOracle is invoked by TreeHist in the final phase (i.e., Final = 1), the collection $\{\tilde{\mathcal{I}}_j : j \in [t]\} = \{\mathcal{I}_j : j \in [t]\}$ and $\gamma = t$. We will use the generic notation $\{\tilde{\mathcal{I}}_j : j \in [t]\}$ and $\gamma$ since the same proof works for both cases.

There are three main sources of randomness. The first is due to randomness in the collection $\{\tilde{\mathcal{I}}_j : j \in [t]\}$ induced by the random partitioning of users via $\Pi$. The second source of randomness is due to the randomness in the choice of the $t$ hash pairs $(h_1, g_1), \ldots, (h_t, g_t)$. The third source of randomness is due to the random row indices $r_i, i \in [n]$, generated by $\mathcal{Q}$, and the randomization for privacy (step 7 in Algorithm 1).

Before we discuss the guarantees we can attain under these sources of randomness, we first introduce some notation. For $j \in [t]$, $v \in \widehat{\mathcal{V}}$, we define $f_{\tilde{\mathcal{I}}_j}(v) \triangleq \sum_{i \in \tilde{\mathcal{I}}_j} \mathbf{1}\left(v_i[1:\ell] = v\right)$; that is, $f_{\tilde{\mathcal{I}}_j}(v)$ is the number of users in $\tilde{\mathcal{I}}_j$ whose the $\ell$-prefix of their items is $v$, and define $n_{\tilde{\mathcal{I}}_j} \triangleq |\tilde{\mathcal{I}}_j|$, i.e., $n_{\tilde{\mathcal{I}}_j}$ is the number of users in $\tilde{\mathcal{I}}_j$. Let $\hat{f}_{\tilde{\mathcal{I}}_1}(v), \ldots, \hat{f}_{\tilde{\mathcal{I}}_t}(v)$ be independent Poisson random variables with mean $\frac{f(v)}{\gamma}$, and let $\hat{n}_{\tilde{\mathcal{I}}_1}, \ldots, \hat{n}_{\tilde{\mathcal{I}}_t}$ be independent Poisson random variables with mean $\frac{n}{\gamma}$.

For each $v \in \widehat{\mathcal{V}}$, let $\mathcal{G}_1(v) = \left\{j \in [t] : |\gamma f_{\tilde{\mathcal{I}}_j}(v) - f(v)| \leq 4\sqrt{n\gamma}, \frac{n}{2} \leq \gamma n_{\tilde{\mathcal{I}}_j} \leq 2n\right\}$.

**Claim 5.4.** *With probability $1 - \frac{\beta}{3 \log d}$ over the randomness in $\{\tilde{\mathcal{I}}_j : j \in [t]\}$, for every $v \in \widehat{\mathcal{V}}$, we have $|\mathcal{G}_1(v)| \geq \frac{9}{10} t$.*

Fix $v \in \widehat{\mathcal{V}}$ and $j \in [t]$. Consider the Poisson random variables $h\hat{f}_{\tilde{\mathcal{I}}_j}(v)$ and $\hat{n}_{\tilde{\mathcal{I}}_j}$ with mean $\frac{f(v)}{\gamma}$ and $\frac{n}{\gamma}$, respectively. By Theorem 2.5 (a tail bound for the Poisson distribution), the union bound, and assuming that $n \geq 48\gamma$ (the assumption stated in the lemma), then with probability at least 0.996, we have

$$|\gamma \hat{f}_{\tilde{\mathcal{I}}_j}(v) - f(v)| \leq 4\sqrt{n\gamma}, \quad \frac{n}{2} \leq \gamma \hat{n}_{\tilde{\mathcal{I}}_j} \leq 2n.$$

Now, consider the sequences of independent Poisson random variables $\hat{f}_{\tilde{\mathcal{I}}_1}(v), \ldots, \hat{f}_{\tilde{\mathcal{I}}_t}(v)$ and $\hat{n}_{\tilde{\mathcal{I}}_1}, \ldots, \hat{n}_{\tilde{\mathcal{I}}_t}$ as defined above. Using the fact that $t \geq 110 \log(n/\beta) \geq 55 \log\left(\frac{3en \log d}{\beta}\right)$, then by Chernoff's bound, with probability at least $1 - \frac{\beta}{3en \log d}$, we have

$$\left|\left\{j \in [t] : |\gamma \hat{f}_{\tilde{\mathcal{I}}_j}(v) - f(v)| \leq 4\sqrt{n\gamma}, \frac{n}{2} \leq \gamma \hat{n}_{\tilde{\mathcal{I}}_j} \leq 2n\right\}\right| \geq \frac{9}{10} t.$$

Using Theorem 2.4 (the Poisson approximation), with probability at least $1 - \frac{\beta}{3\sqrt{n}\log(d)}$, we have $|\mathcal{G}_1(v)| \geq \frac{9}{10} t$. Hence, by the fact that $|\widehat{\mathcal{V}}| \leq \sqrt{n}$ and the union bound, with probability at least $1 - \frac{\beta}{3 \log d}$, for all $v \in \widehat{\mathcal{V}}$, we have $|\mathcal{G}_1(v)| \geq \frac{9}{10} t$.

For the remainder of the proof, we will condition on the event in Claim 5.4.

Let $\mathbf{w}_c$ denote the $c$-th column of the basis matrix $\mathbf{W}$. Since the prefix length $\ell$ is fixed, we will denote $v_i[1:\ell]$ (the $\ell$-prefix of the item of user $i$) as $v_i$ for brevity.

For each $v \in \widehat{\mathcal{V}}$, let $\mathcal{G}_2(v) = \left\{j \in [t] : \left|\gamma \sum_{i \in \tilde{\mathcal{I}}_j} \frac{g_j(v_i) \cdot g_j(v)}{m} \cdot \langle \mathbf{w}_{h_j(v_i)}, \mathbf{w}_{h_j(v)}\rangle - f(v)\right| \leq 8\sqrt{\gamma n}\right\}$

**Claim 5.5.** *Conditioned on the event in Claim 5.4, with probability at least $1 - \frac{\beta}{3 \log d}$ over the choice of the $t$ hash pairs $\{(h_j, g_j) : j \in [t]\}$, for all $v \in \widehat{\mathcal{V}}$, we have $|\mathcal{G}_2(v)| \geq \frac{4}{5} t$.*

Fix $v \in \widehat{\mathcal{V}}$ and $j \in \mathcal{G}_1(v)$. First consider $\left|\sum_{i \in \tilde{\mathcal{I}}_j} \frac{g_j(v_i) \cdot g_j(v)}{m} \cdot \langle \mathbf{w}_{h_j(v_i)}, \mathbf{w}_{h_j(v)}\rangle - f_{\tilde{\mathcal{I}}_j}(v)\right|$. Note that the columns of $\mathbf{W}$ are orthogonal, and each of them has norm $\sqrt{m}$. Hence, the error quantity above comes from those $i \in \tilde{\mathcal{I}}_j$ with $v_i \neq v$ and yet $h_j(v_i) = h_j(v)$. In particular, we can write this error as

$$\left|\sum_{i \in \tilde{\mathcal{I}}_j} z_i\, g_j(v_i)\, g_j(v)\right|$$

where $z_i = \mathbf{1}\left(v_i \neq v, h_j(v_i) = h_j(v)\right)$. Define $\mathcal{V}_{\text{Bad}} = \{\hat{v} \in \widehat{\mathcal{V}} : f_{\tilde{\mathcal{I}}_j}(\hat{v}) \geq 2\sqrt{n_{\tilde{\mathcal{I}}_j}}\}$. Note that $\left|\mathcal{V}_{\text{Bad}}\right| \leq \sqrt{n_{\tilde{\mathcal{I}}_j}}/2$. Now, observe that

$$\mathop{\mathbb{P}}_{h_j, g_j}\left[\left|\sum_{i \in \tilde{\mathcal{I}}_j} z_i\, g_j(v_i)\, g_j(v)\right| > 2\sqrt{n_{\tilde{\mathcal{I}}_j}}\right] \leq \mathop{\mathbb{P}}_{h_j, g_j}\left[\exists\, \hat{v} \in \mathcal{V}_{\text{Bad}} : h_j(\hat{v}) = h_j(v)\right]$$

$$+ \mathop{\mathbb{P}}_{h_j, g_j}\left[\left|\sum_{i \in \tilde{\mathcal{I}}_j} z_i\, g_j(v_i)\, g_j(v)\right| > 2\sqrt{n_{\tilde{\mathcal{I}}_j}}\ \Big|\ \forall\, \hat{v} \in \mathcal{V}_{\text{Bad}}, h_j(\hat{v}) \neq h_j(v)\right]$$

By the pairwise independence property of $h_j$ and the union bound, the first probability term on the right hand side is bounded from above by $\frac{|\mathcal{V}_{\text{Bad}}|}{m} = \frac{\sqrt{n_{\tilde{\mathcal{I}}_j}}}{2m}$. We now consider the second probability term. Note that the event we conditioned on in the second probability term implies that for every $i \in \tilde{\mathcal{I}}_j$ where $v_i \neq v$, we must have $f_{\tilde{\mathcal{I}}_j}(v_i) < 2\sqrt{n_{\tilde{\mathcal{I}}_j}}$. Hence, conditioned on this event, by the pairwise independence of each of $h_j$ and $g_j$, we have

$$\mathbf{Var}\left[\sum_{i \in \tilde{\mathcal{I}}_j} z_i\, g_j(v_i)\, g_j(v)\right] = \sum_{i \in \tilde{\mathcal{I}}_j} \mathbf{Var}\left[z_i\, g_j(v_i)\right] + \sum_{\substack{i,k \in \tilde{\mathcal{I}}_j:\\ v_i = v_k \neq v}} \mathbb{E}\left[z_i z_k\right]\mathbb{E}\left[g_j(v_i) g_j(v_k)\right]$$

$$\leq n_{\tilde{\mathcal{I}}_j}/m + 2n_{\tilde{\mathcal{I}}_j}^{3/2}/m \leq 3n_{\tilde{\mathcal{I}}_j}^{3/2}/m.$$

Hence, by using Chebyshev's inequality, the second probability term is bounded by $\frac{3\sqrt{n_{\tilde{\mathcal{I}}_j}}}{4m}$. Hence, we have

$$\mathop{\mathbb{P}}_{h_j, g_j}\left[\left|\sum_{i \in \tilde{\mathcal{I}}_j} z_i\, g_j(v_i)\, g_j(v)\right| > 2\sqrt{n_{\tilde{\mathcal{I}}_j}}\right] \leq \frac{5\sqrt{n_{\tilde{\mathcal{I}}_j}}}{4m} \leq \frac{\sqrt{\log(n/\beta)}}{25\sqrt{\gamma}} \leq \frac{1}{250},$$

where the last inequality follows from the fact that $j \in \mathcal{G}_1(v)$ and the fact that $m \geq 48\sqrt{\frac{n}{\log(n/\beta)}}$. Thus, with probability at least 0.996, for every $v \in \widehat{\mathcal{V}}$ and every $j \in \mathcal{G}_1(v)$, we have

$$\left|\sum_{i \in \tilde{\mathcal{I}}_j} \frac{g_j(v_i) \cdot g_j(v)}{m} \cdot \langle \mathbf{w}_{h_j(v_i)}, \mathbf{w}_{h_j(v)}\rangle - f_{\tilde{\mathcal{I}}_j}(v)\right| \leq 2\sqrt{n_{\tilde{\mathcal{I}}_j}}.$$

Now, as we conditioned on the event in Claim 5.4, this implies that with probability at least 0.996, for every $v \in \widehat{\mathcal{V}}$ and every $j \in \mathcal{G}_1(v)$, we have

$$\left|\gamma \sum_{i \in \tilde{\mathcal{I}}_j} \frac{g_j(v_i) \cdot g_j(v)}{m} \cdot \langle \mathbf{w}_{h_j(v_i)}, \mathbf{w}_{h_j(v)}\rangle - f(v)\right|$$

$$\leq \left|\gamma \sum_{i \in \tilde{\mathcal{I}}_j} \frac{g_j(v_i) \cdot g_j(v)}{m} \cdot \langle \mathbf{w}_{h_j(v_i)}, \mathbf{w}_{h_j(v)}\rangle - \gamma f_{\tilde{\mathcal{I}}_j}(v)\right| + \left|\gamma f_{\tilde{\mathcal{I}}_j}(v) - f(v)\right|$$

$$\leq 8\sqrt{\gamma n}.$$

Conditioned on the event of Claim 5.4, $|\mathcal{G}_1(v)| \geq \frac{9}{10}t \geq 99\log(n/\beta) \geq 49\log\left(\frac{3\sqrt{n}\log d}{\beta}\right)$. We note also that, for all $j \in [t]$, the above sums are independent. Thus, conditioned on the event of Claim 5.4, by Chernoff's bound, with probability at least $1 - \frac{\beta}{3\sqrt{n}\log d}$, we have $|\mathcal{G}_2(v)| \geq \frac{22}{25}|\mathcal{G}_1(v)| \geq \frac{4}{5}t$. Hence, by the fact that $|\widehat{\mathcal{V}}| \leq \sqrt{n}$ and the

union bound, with probability at least $1 - \frac{\beta}{3\log d}$, for all $v \in \widehat{\mathcal{V}}$, we have $|\mathcal{G}_2(v)| \geq \frac{4}{5}t$. We continue the proof while conditioning on this event as well.

For every $i \in [n]$ let $r_i \leftarrow [m]$ be the row index chosen uniformly at random for user $i$ using the random function $\mathcal{Q}$, and let $y_i$ denote the randomized bit generated by user $i$ in step 7 of Algorithm 1. As was denoted in algorithm FreqOracle, for each $v \in \widehat{\mathcal{V}}$, let $\hat{f}_j(v) = \gamma a_\epsilon \sum_{i \in \tilde{\mathcal{I}}_j} y_i \, g_j(v) \, W_{r_i, h_j(v)}$ denote the $j$-th frequency estimate of FreqOracle for $v \in \widehat{\mathcal{V}}$.

For each $v \in \widehat{\mathcal{V}}$, let
$$\mathcal{G}_3(v) = \left\{ j \in [t] : |\hat{f}_j(v) - f(v)| \leq 14\sqrt{n\gamma}/\epsilon \right\}.$$

**Claim 5.6.** *Conditioned on the events in Claims 5.4 and 5.5, with probability at least $1 - \frac{\beta}{3\log d}$ over the randomness in $\{(r_i, y_i) : i \in \tilde{\mathcal{I}}_j, j \in [t]\}$, for all $v \in \widehat{\mathcal{V}}$, we have $|\mathcal{G}_3(v)| \geq \frac{7}{10}t$*

Fix $v \in \widehat{\mathcal{V}}$ and $j \in \mathcal{G}_2(v)$. Note that, conditioned on any realization for $\tilde{\mathcal{I}}_j$, each term in the sum
$$\sum_{i \in \tilde{\mathcal{I}}_j} \left( a_\epsilon \, y_i \, g_j(v) \, W_{r_i, h_j(v)} - \frac{g_j(v_i) \, g_j(v)}{m} \langle \mathbf{w}_{h_j(v_i)}, \mathbf{w}_{h_j(v)} \rangle \right)$$

is independent, zero mean random variable whose support length is bounded by $a_\epsilon + 1 = O\left(\frac{1}{\epsilon}\right)$. Hence, by Chernoff's bound, with probability at least 0.99, we have
$$\left| \sum_{i \in \tilde{\mathcal{I}}_j} \left( a_\epsilon \, y_i \, g_j(v) \, W_{r_i, h_j(v)} - \frac{g_j(v_i) \, g_j(v)}{m} \langle \mathbf{w}_{h_j(v_i)}, \mathbf{w}_{h_j(v)} \rangle \right) \right| \leq 4\sqrt{n_{\tilde{\mathcal{I}}_j}}/\epsilon.$$

Thus, conditioned on the events in Claims 5.4 and 5.5, with probability at least 0.99, we have
$$\left| \gamma \sum_{i \in \tilde{\mathcal{I}}_j} a_\epsilon \, y_i \, g_j(v) \, W_{r_i, h_j(v)} - f(v) \right| \leq 14\sqrt{\gamma n}/\epsilon,$$

i.e., $|\hat{f}_j(v) - f(v)| \leq 14\sqrt{n\gamma}/\epsilon$.

Conditioned on the event of Claim 5.5, $|\mathcal{G}_2(v)| \geq \frac{4}{5}t \geq 88\log(n/\beta) \geq 44\log\left(\frac{3\sqrt{n}\log d}{\beta}\right)$. We note also that the above sums for $j = 1, \ldots, t$ are independent. Thus, conditioned on the event of Claim 5.5, by Chernoff's bound, with probability at least $1 - \frac{\beta}{3\sqrt{n}\log d}$, we have $|\mathcal{G}_3(v)| \geq \frac{22}{25}|\mathcal{G}_2(v)| \geq \frac{88}{125}t \geq \frac{7}{10}t$. Hence, by the fact that $|\widehat{\mathcal{V}}| \leq \sqrt{n}$ and the union bound, with probability at least $1 - \frac{\beta}{3\log d}$, for all $v \in \widehat{\mathcal{V}}$, we have $|\mathcal{G}_3(v)| \geq \frac{7}{10}t$.

By combining Claims 5.4, 5.5, and 5.6, we conclude that with probability at least $1 - \frac{\beta}{\log d}$, for all $v \in \widehat{\mathcal{V}}$, we have
$$\left| \{ j \in [t] : |\hat{f}_j(v) - f(v)| \leq 14\sqrt{n\gamma}/\epsilon \} \right| > \frac{1}{2}t.$$

Since for any item $v$, the final frequency estimate $\hat{f}(v)$ generated by FreqOracle is the median of $\hat{f}_1(v), \ldots, \hat{f}_t(v)$, then the above implies that with probability at least $1 - \frac{\beta}{\log d}$, for all $v \in \widehat{\mathcal{V}}$, we have $|\hat{f}(v) - f(v)| \leq 14\sqrt{n\gamma}/\epsilon$. This completes the proof of the lemma. $\square$

# 6 Locally Private Heavy-hitters bit-by-bit: The `Bitstogram` Protocol

We will use the following notation. Let $S \in \mathcal{V}^n$ be a database, which may be distributed across $n$ users (each holding one row). For $v \in \mathcal{V}$, we will be interested in estimating the the duplicity of $v$ in $S$, i.e., $f_S(v) = |\{v_i \in S : v_i = v\}|$.

## 6.1 Warmup: A Simple Protocol for Heavy-Hitters

For readability, we first present a simplification of our protocol that captures most of the ideas. We will later modify the construction in order to reduce the worst-case error, space complexity, and time complexity of the protocol.

### 6.1.1 Frequency Oracle

Our protocols use the simple local randomizer $\mathcal{R}$ (Algorithm 4), where every user holds one bit, and flips it with probability $1/(e^\epsilon + 1)$.

---
**Algorithm 4** $\mathcal{R}$: Basic Randomizer

**Inputs:** $x \in \{\pm 1\}$, and privacy parameter $\epsilon$.

1. Generate and return a random bit $z = \begin{cases} x & \text{w.p. } e^\epsilon/(e^\epsilon + 1) \\ -x & \text{w.p. } 1/(e^\epsilon + 1) \end{cases}$

---

---
**Algorithm 5** `ExplicitHist`

**Public randomness:** Uniformly random matrix $Z \in \{\pm 1\}^{d \times n}$.

**Setting:** Each player $j \in [n]$ holds a value $v_j \in \mathcal{V}$. Define $S = (v_1, \cdots, v_n)$.
Define $\widetilde{S} = (\tilde{v}_1, \cdots, \tilde{v}_n)$ where $\tilde{v}_j = Z[v_j, j]$.

**Oracle:** LR Oracle access to $\widetilde{S}$.

1. For $j \in [n]$ let $y_j \leftarrow LR_{\widetilde{S}}(j, \mathcal{R})$.
2. On input $v \in \mathcal{V}$, return $a(v) = \frac{e^\epsilon + 1}{e^\epsilon - 1} \cdot \sum_{j \in [n]} y_j \cdot Z[v, j]$, and wait for the next input.

---

**Lemma 6.1.** *Let $\epsilon \leq 1$, and fix a subset $V \subseteq \mathcal{V}$ of size $d' \leq d$. With probability at least $1 - \beta$, algorithm `ExplicitHist` answers every $v \in V$ with $a(v)$ satisfying:*

$$|a(v) - f_S(v)| \leq \frac{3}{\epsilon}\sqrt{n \cdot \ln(4d'/\beta)}.$$

*Proof.* Fix $v \in V$, and denote $c(v) = \sum_{j \in [n]} y_j \cdot Z[v, j]$, and recall that algorithm `ExplicitHist` answers the query $v$ with $a(v) = \frac{e^\epsilon + 1}{e^\epsilon - 1} \cdot c(v)$. We start by analyzing the expectation of $c(v)$:

$$\begin{aligned}
\mathbb{E}[c(v)] &= \sum_{j \in [n]} \mathbb{E}[y_j \cdot Z[v, j]] = \sum_{j \in [n]: v_j = v} \mathbb{E}[y_j \cdot Z[v, j]] + \sum_{j \in [n]: v_j \neq v} \mathbb{E}[y_j \cdot Z[v, j]] \\
&= \sum_{j \in [n]: v_j = v} \mathbb{E}[y_j \cdot Z[v, j]] + \sum_{j \in [n]: v_j \neq v} \mathbb{E}[y_j] \cdot \mathbb{E}[Z[v, j]] = f_S(v) \cdot \frac{e^\epsilon - 1}{e^\epsilon + 1}.
\end{aligned}$$

That is, $c(v)$ can be expressed as two sums of $\pm 1$ independent random variables: $f_S(v)$ variables with expectation $\frac{e^\epsilon - 1}{e^\epsilon + 1}$, and $(n - f_S(v))$ variables with expectation 0. Using the Hoeffding bound, with probability at least $1 - \frac{\beta}{d'}$ we have that $|c(v) - \frac{e^\epsilon - 1}{e^\epsilon + 1} \cdot f_S(v)| \leq \sqrt{n \cdot \ln(4d'/\beta)}$. That is, $|a(v) - f_S(v)| \leq \frac{e^\epsilon + 1}{e^\epsilon - 1} \cdot \sqrt{n \cdot \ln(4d'/\beta)}$. Using the union bound, this holds simultaneously for every $v \in V$ with probability at least $1 - \beta$. □

**Observation 6.1.** For the analysis above it suffices that, for every $j \in [n]$, the entries of column $j$ of $Z$ are only pairwise independent. Furthermore, appealing to Lemma 2.3 (concentration of $k$-wise independent random variables) instead of the Hoeffding bound, is suffices that, for every $v \in \mathcal{V}$, the entries of row $v$ of $Z$ are only $k$-wise independent, for $k = 3 \ln(d/\beta)$.

### 6.1.2 A Simple Heavy Hitters Protocol

---
**Algorithm 6** `SuccinctHist`

**Public randomness:** Random hash function $h : \mathcal{V} \to [T]$.
$\qquad\qquad\qquad\qquad$ Random partition of $[n]$ into $\log d$ subsets $I_1, \cdots, I_{\log d}$.

**Setting:** Each player $j \in [n]$ holds a value $v_j \in \mathcal{V}$. Define $S = (v_1, \cdots, v_n)$.
$\qquad\qquad$ For $\ell \in [\log d]$, let $S_\ell = (h(v_j), v_{j,\ell})_{j \in I_\ell}$, where $v_{j,\ell}$ is bit $\ell$ of $v_j$.
$\qquad\qquad$ That is, $S_\ell$ is a database over the domain $[T] \times \{0, 1\}$.

1. For $\ell \in [\log d]$, use `ExplicitHist`$(S_\ell)$ with $\frac{\epsilon}{2}$ to get $a_\ell(t, b)$ for all $(t, b) \in [T] \times \{0, 1\}$.
2. For $t \in [T]$, define $\hat{v}_t \in \mathcal{V}$, where bit $\ell$ of $\hat{v}_t$ is $\hat{v}_{t,\ell} = \operatorname{argmax}\{a_\ell(t, 0), a_\ell(t, 1)\}$.
3. Use `ExplicitHist`$(S)$ with privacy parameter $\frac{\epsilon}{2}$ to obtain $a(\hat{v}_t)$ for all $t \in [T]$.
4. Return list $L = \{(\hat{v}_t, a(\hat{v}_t)) : t \in [T]\}$.

---

**Lemma 6.2.** *Let $\epsilon \leq 1$, denote $w \triangleq 32 \log(d) \log(16 \log d) + \frac{48}{\epsilon} \sqrt{2n \log d \cdot \ln(64 \log d)}$, and set $T = \frac{32n}{w}$. Algorithm `SuccinctHist` returns a list $L$ of length $T$ satisfying:*

1. *With probability $1 - \beta$, for every $(v, a) \in L$ we have that $|a - f_S(v)| \leq \frac{6}{\epsilon} \sqrt{n \cdot \ln(4T/\beta)}$.*
2. *For every $v \in \mathcal{V}$ s.t. $f_S(v) \geq w$, with probability $1/2$ we have that $v$ is in $L$.*

**Remark 6.2.** The $\log \log d$ factors in the above lemma can be removed by using an error correction code s.t. in order to recover a "heavy-hitter" $v^*$ it suffices to recover correctly only part of its (encoded) bits.

*Proof.* Item 1 of the lemma follows directly from Lemma 6.1. We now prove item 2. Assuming that $n \geq 12 \log(d) \log(12 \log d)$, by the Chernoff bound, with probability at least $7/8$ (over partitioning $[n]$ into subsets $I_1, \cdots, I_{\log d}$), for every $\ell \in [\log d]$ we have that $\frac{n}{2 \log d} \leq |I_\ell| \leq \frac{2n}{\log d}$. We continue the analysis assuming that this is the case.

Fix $v^* \in \mathcal{V}$ s.t. $f_S(v^*) \geq w$, and consider the following good event (over sampling $h$):

$$\textbf{Event } E_1: \ |\{v \in S : v \neq v^* \text{ and } h(v) = h(v^*)\}| \leq w/4.$$

Event $E_1$ states that $v^*$ is mapped (by the hash function $h$) into a cell without too many collisions with different input elements. Denote $t^* = h(v^*)$. While the duplicity of $v^*$ in $S$ is at least $w$, event $E_1$ states that other than $v^*$ there are at most $w/4$ elements which are mapped into $t^*$. That is, $v^*$ dominates the cell $t^*$. We first show that if $E_1$ occurs, then w.h.p. $v^*$ is in the list $L$.

Asserting that $w \geq 32 \log(d) \log(16 \log d)$, by the Chernoff bound we get that with probability $7/8$ (over partitioning $[n]$ into subsets $I_1, \cdots, I_{\log d}$), for every $\ell \in [\log d]$ we have that

$$f_{S_\ell}(h(v^*), v_\ell^*) \geq f_{S_\ell}(h(v^*), 1 - v_\ell^*) + \frac{w}{4 \log d}.$$

If that is the case, then by the properties of algorithm `ExplicitHist`, for $w \geq \frac{48}{\epsilon} \sqrt{2n \log d \cdot \ln(64 \log d)}$, with probability at least $1 - \frac{1}{8 \log d}$ we have that $\hat{v}_{t^*, \ell} = v_\ell^*$, where $t^* = h(v^*)$. Using the union bound, this holds simultaneously for all $\ell \in [\log d]$ with probability at least $7/8$, in which case $v^* = \hat{v}_{t^*}$ is in the list $L$.

It remains to show that Event $E_1$ occurs with high probability. To that end, observe that

$$\mathbb{E}_h \left[ |\{v \in S : v \neq v^* \text{ and } h(v) = h(v^*)\}| \right] = \sum_{v \in S : v \neq v^*} \mathbb{E}_h \left[ \mathbb{1}_{h(v) = h(v^*)} \right] \leq \frac{n}{T}.$$

Thus, by Markov's inequality, we have that

$$\Pr\left[|\{v \in S : v \neq v^* \text{ and } h(v) = h(v^*)\}| \geq \frac{8n}{T}\right] \leq \frac{1}{8}.$$

Setting $T = \frac{32n}{w}$ completes the proof. $\square$

## 6.2 Reducing Space and Time Complexities

### 6.2.1 Space and Time Efficient Frequency Oracle

---
**Algorithm 7** Hashtogram
---
**Public randomness:** Random partition of $[n]$ into $R$ subsets $I_1, \cdots, I_R$.
    Random hash functions $h_1, \cdots, h_R$ mapping $\mathcal{V}$ to $[T]$.
    Uniformly random matrix $Z \in \{\pm 1\}^{T \times n}$.

**Setting:** Each player $j \in [n]$ holds a value $v_j \in \mathcal{V}$. Define $S = (v_1, \cdots, v_n)$.
    Define $\widetilde{S} = (\tilde{v}_1, \cdots, \tilde{v}_n)$, where $\tilde{v}_j = Z[h_r(v_j), j]$ and $h_r$ is s.t. $j \in I_r$.

**Oracle:** LR Oracle access to $\widetilde{S}$.

1. For $j \in [n]$ let $y_j \leftarrow LR_{\widetilde{S}}(j, \mathcal{R})$.

2. For every $(r, t) \in [R] \times [T]$ compute $a_r(t) = \frac{e^\epsilon + 1}{e^\epsilon - 1} \cdot \sum_{j \in I_r} y_j \cdot Z[t, j]$.

3. On input $v \in \mathcal{V}$, return $a(v) = R \cdot \text{Median}\{a_1(h_1(v)), \cdots, a_R(h_R(v))\}$, and wait for the next input.

---

**Lemma 6.3.** *Algorithm* Hashtogram *satisfies $\epsilon$-LDP.*

**Lemma 6.4.** *Let $\epsilon \leq 1$, and fix a subset $V \subseteq \mathcal{V}$ of size $d' \leq d$ to be queried to algorithm* Hashtogram*. Let algorithm* Hashtogram *be executed with $R \geq 132 \log(4d'/\beta)$, and $T \geq \epsilon^2 \cdot \log(d'/\beta) + \epsilon \cdot \sqrt{n/\log(d'/\beta)}$, and $n \geq 8R \log(8d'/\beta)$. With probability at least $1 - \beta$, algorithm* Hashtogram *answers every $v \in V$ with $a(v)$ satisfying:*

$$|a(v) - f_S(v)| \leq \frac{27}{\epsilon} \cdot \sqrt{nR \log(\frac{2Rd'}{\beta})}.$$

Observe that the error in the lemma is sub-optimal, as the optimal error behaves like $\frac{1}{\epsilon}\sqrt{n \log d}$. However, we will only use Lemma 6.4 with constant $d'$ and constant $\beta$, and hence, will not be effected by this issue. A similar analysis (for the same algorithm) gives better bounds for other settings of parameters. Specifically,

**Lemma 6.5.** *Let $\epsilon \leq 1$. Fix a subset $V \subseteq \mathcal{V}$ of size $d' \leq d$ to be queried to algorithm* Hashtogram*. Let algorithm* Hashtogram *be executed with $R \geq 300 \log(12nd'/\beta)$ and $n \geq 43R$ and $T \geq \epsilon \cdot \sqrt{n/\log(nd'/\beta)}$. With probability at least $1 - \beta$, algorithm* Hashtogram *answers every $v \in V$ with $a(v)$ satisfying:*

$$|a(v) - f_S(v)| \leq \frac{400}{\epsilon} \cdot \sqrt{n \log\left(\frac{12nd'}{\beta}\right)}.$$

As the analysis of the two lemmas are very similar, we only present the proof of Lemma 6.5. The proof of Lemma 6.4 appears in Section B for completness.

*Proof of Lemma 6.5.* Consider the following good event:

**Event $E_1$ (over sampling $h_1, \cdots, h_R$):**
For every query $v^* \in V$ there exists a subset $R_1^{v^*} \subseteq [R]$ of size $|R_1^{v^*}| \geq \frac{7}{8}R$ s.t. for every $r^* \in R_1^{v^*}$ it holds that
$|\{v \in S : v \neq v^*$ and $h_{r^*}(v) = h_{r^*}(v^*)\}| \leq \frac{16n}{T}$.

---

Event $E_1$ states that for at least $7R/8$ of the hash functions, we have that $v^*$ is mapped into a cell without too many collisions with different input elements. Informally, for every single hash function $h_r$, algorithm `Hashtogram` estimates the number of occurrences of $h_r(v^*)$ in $S$. Hence, if event $E_1$ occurs, then most of the estimations result in accurate answers. We start by showing that event $E_1$ happens with high probability. To that end, fix $v^* \in V$ and fix $r^* \in [R]$. We have that

$$\mathbb{E}_{h_{r^*}}\left[|\{v \in S : v \neq v^* \text{ and } h_{r^*}(v) = h_{r^*}(v^*)\}|\right] = \sum_{v \in S: v \neq v^*} \mathbb{E}_{h_{r^*}}\left[\mathbb{1}_{h_{r^*}(v) = h_{r^*}(v^*)}\right] \leq \frac{n}{T}.$$

Thus, by Markov's inequality, we have that

$$\Pr_{h_{r^*}}\left[|\{v \in S : v \neq v^* \text{ and } h_{r^*}(v) = h_{r^*}(v^*)\}| \geq \frac{16n}{T}\right] \leq \frac{1}{16}.$$

As the hash functions are independent from each other, for $R \geq 48\ln(\frac{d'}{\beta})$, by the Chernoff bound we get that with probability at least $1 - \beta/d'$ (over sampling $h_1, \cdots, h_R$) there exists a subset $R_1^{v^*} \subseteq [R]$ of size $|R_1^{v^*}| \geq \frac{7}{8}R$ s.t. for every $r^* \in R_1^{v^*}$ it holds that

$$|\{v \in S : v \neq v^* \text{ and } h_{r^*}(v) = h_{r^*}(v^*)\}| \leq \frac{16n}{T}.$$

Using the union bound over every $v^* \in V$, we have that event $E_1$ happens with probability at least $1 - \beta$. We continue the analysis assuming that event $E_1$ occurs.

---

**Event $E_2$ (over partitioning $[n]$ into $I_1, \cdots, I_R$):**
There exists a subset $R_2 \subseteq [R]$ of size $|R_2| \geq \frac{7}{8}R$ s.t. for every $r \in R_2$ it holds that $\frac{n}{2R} \leq |I_r| \leq \frac{2n}{R}$.

---

Following Theorem 2.4 (the Poisson approximation), we analyze event $E_2$ in the Poisson case. To that end, let $\hat{I}_1, \cdots, \hat{I}_R$ be independent Poisson random variables with mean $n/R$. Now fix $r \in R$. Using a tail bound for the Poisson distribution (see Theorem 2.5), assuming that $n \geq 43R$ we have that $\Pr[\frac{n}{2R} \leq \hat{I}_r \leq \frac{2n}{r}] \geq \frac{99}{100}$. As $\hat{I}_1, \cdots, \hat{I}_R$ are independent, assuming that $R \geq 300\ln(\frac{3n}{\beta})$, by the Chernoff bound we get that event $E_2$ happens with probability at least $1 - \frac{\beta}{3n}$ in the Poisson case. Hence, by Theorem 2.4, event $E_2$ happens with probability at least $1 - \beta$. We continue the analysis assuming that this is the case.

For every $r \in [R]$, let $S_r = (v_j)_{j \in I_r}$ denote a database containing the data of all users $j$ s.t. $j \in I_r$. Also for $v^* \in V$ and $r \in [R]$ denote $S^{r,v^*} \triangleq \{v \in S : h_r(v) = h_r(v^*)\}$. That is, $|S^{r,v^*}|$ is the number of users $j$ s.t. $h_r(v_j) = h_r(v^*)$. Furthermore, for $v^* \in V$ and $r \in [R]$ denote $I_r^{v^*} \triangleq \{v \in S_r : h_r(v) = h_r(v^*)\}$. That is, $|I_r^{v^*}|$ is the number of users $j$ s.t. $j \in I_r$ and $h_r(v_j) = h_r(v^*)$. Observe that $|S^{r,v^*}| \geq f_S(v^*)$ and that $|I_r^{v^*}| \geq f_{S_r}(v^*)$.

---

**Event $E_3$ (over partitioning $[n]$ into $I_1, \cdots, I_R$):**
For every query $v^* \in V$ there exists a subset $R_3^{v^*} \subseteq [R]$ of size $|R_3^{v^*}| \geq \frac{9}{10}R$ s.t. for every $r^* \in R_3^{v^*}$ it holds that
$\left|R \cdot |I_r^{v^*}| - |S^{r,v^*}|\right| \leq \sqrt{8Rn}$

---

We analyze event $E_3$ in the Poisson case. To that end, fix $v^* \in V$, and let $\hat{I}_1^{v^*}, \cdots, \hat{I}_R^{v^*}$ be independent Poisson random variables with mean $\frac{|S^{r,v^*}|}{R}$. Now fix $r \in [R]$. Using a tail bound for the Poisson distribution (see Theorem 2.5), with probability at least $19/20$ we have that

$$\left| R \cdot \hat{I}_r^{v^*} - |S^{r,v^*}| \right| \leq \sqrt{8Rn}. \tag{1}$$

As $\hat{I}_1^{v^*}, \cdots, \hat{I}_R^{v^*}$ are independent, assuming that $R \geq 300 \ln(\frac{3nd'}{\beta})$, by the Chernoff bound we get that with probability at least $1 - \frac{\beta}{3nd'}$, Inequality (1) holds for at least $18/20$ choices of $r \in [R]$. By Theorem 2.4 (the Poisson approximation), with probability at least $1 - \frac{\beta}{d'}$, this is also the case for the random variables $|I_r^{v^*}|$. That is, with probability at least $1 - \frac{\beta}{d'}$, there exists a subset $R_3^{v^*} \subseteq [R]$ of size $|R_3^{v^*}| \geq \frac{9}{10}R$ s.t. for every $r^* \in R_3^{v^*}$ it holds that

$$\left| R \cdot |I_r^{v^*}| - |S^{r,v^*}| \right| \leq \sqrt{8Rn}.$$

Using the union bound over every choice of $v^* \in V$, we get that event $E_3$ happens with probability at least $1 - \beta$. We continue with the analysis assuming that this is the case.

---

**Event $E_4$ (over sampling $Z$ and the coins of the local randomizers):**
For every query $v^* \in V$ there exists a subset $R_4^{v^*} \subseteq [R]$ of size $|R_4^{v^*}| \geq \frac{7}{8} \cdot \frac{9}{10}R$ s.t. for every $r^* \in R_4^{v^*}$ it holds that $\left| R \cdot a_{r^*}(h_{r^*}(v^*)) - R \cdot |I_{r^*}^{v^*}| \right| \leq \frac{e^\epsilon + 1}{e^\epsilon - 1} \cdot \sqrt{11nR}$.

---

For $v^* \in V$ and $r \in [R]$ denote $c_r(v^*) = \sum_{j \in I_r} y_j \cdot Z[h_r(v^*), j]$, and recall that algorithm Hashtogram answers the query $v^* \in \mathcal{V}$ with $a(v^*) = R \cdot \frac{e^\epsilon + 1}{e^\epsilon - 1} \cdot \text{Median}\{c_r(v^*)\}_{r \in [R]}$. Fix $v^* \in V$ and $r \in R_2$ (where $R_2 \subseteq [R]$ is the subset from event $E_2$). We now analyze the expectation of $c_r(v^*)$:

$$\begin{aligned}
\mathbb{E}[c_r(v^*)] &= \sum_{j \in I_r} \mathbb{E}\left[y_j \cdot Z[h_r(v^*), j]\right] \\
&= \sum_{j \in I_r:\, h_r(v_j) = h_r(v^*)} \mathbb{E}\left[y_j \cdot Z[h_r(v^*), j]\right] + \sum_{j \in I_r:\, h_r(v_j) \neq h_r(v^*)} \mathbb{E}\left[y_j \cdot Z[h_r(v^*), j]\right] \\
&= \sum_{j \in I_r:\, h_r(v_j) = h_r(v^*)} \mathbb{E}\left[y_j \cdot Z[h_r(v^*), j]\right] + \sum_{j \in I_r:\, h_r(v_j) \neq h_r(v^*)} \mathbb{E}\left[y_j\right] \cdot \mathbb{E}\left[Z[h_r(v^*), j]\right] \\
&= |\{v \in S_r : h_r(v) = h_r(v^*)\}| \cdot \frac{e^\epsilon - 1}{e^\epsilon + 1} \triangleq |I_r^{v^*}| \cdot \frac{e^\epsilon - 1}{e^\epsilon + 1}
\end{aligned}$$

That is, $c_r(v^*)$ can be expressed as two sums of $\pm 1$ independent random variables: $|I_r^{v^*}|$ variables with expectation $\frac{e^\epsilon - 1}{e^\epsilon + 1}$, and $(|I_r| - |I_r^{v^*}|)$ variables with expectation $0$ (recall that by event $E_2$ we have $\frac{n}{2R} \leq |I_r| \leq \frac{2n}{R}$). Using the Hoeffding bound, with probability at least $43/44$ we have that $\left| c_r(v^*) - \frac{e^\epsilon - 1}{e^\epsilon + 1} \cdot |I_r^{v^*}| \right| \leq \sqrt{11n/R}$. That is,

$$\left| R \cdot a_r(h_r(v^*)) - R \cdot |I_r^{v^*}| \right| \leq \frac{e^\epsilon + 1}{e^\epsilon - 1} \cdot \sqrt{11nR}. \tag{2}$$

Fix $v^* \in V$, and observe that the above sums are independent for different values of $r$. Hence, using the Chernoff bound and asserting that $R \geq 150 \ln(d'/\beta)$, for that fixed $v^* \in M$, with probability at least $1 - \beta/d'$ we have that Inequality (2) holds for at least $7R/8$ choices of $r \in R_1$. Using the union bound, with probability at least $1 - \beta$, this is true for every $v^* \in V$ simultaneously. That is, event $E_4$ happens with probability at least $1 - \beta$. We continue the analysis assuming that event $E_4$ occurs.

We are now ready to complete the proof. Fix $v^* \in V$. Combining events $E_3$ and $E_4$, we get that for every $r \in R_3^{v^*} \cap R_4^{v^*}$

$$\left| R \cdot a_r(h_r(v^*)) - |S^{r,v^*}| \right| \leq \frac{e^\epsilon + 1}{e^\epsilon - 1} \cdot \sqrt{11nR} + \sqrt{8Rn}. \tag{3}$$

Recall that for every $r \in [R]$ we have that $|S^{r,v^*}| \geq f_S(v^*)$. Furthermore, by event $E_1$, for every $r \in R_1^{v^*}$ we have that $|S^{r,v^*}| \leq f_S(v^*) + \frac{16n}{T}$. Hence, for every $r \in R_1^{v^*} \cap R_3^{v^*} \cap R_4^{v^*}$ we have that

$$\left| R \cdot a_r(h_r(v^*)) - f_S(v^*) \right| \leq \frac{e^\epsilon + 1}{e^\epsilon - 1} \cdot \sqrt{11nR} + \sqrt{8Rn} + \frac{16n}{T} \triangleq \mathrm{error}(v^*).$$

That is, for every $r \in R_1^{v^*} \cap R_3^{v^*} \cap R_4^{v^*}$ we have that $R \cdot a_r(h_r(v^*))$ is accurate up to $\mathrm{error}(v^*)$. As $|R_1^{v^*} \cap R_3^{v^*} \cap R_4^{v^*}| \geq \frac{9}{16}R$, and as algorithm Hashtogram answers $v^*$ with $a(v^*)$ chosen as the median of $\{R \cdot a_r(h_r(v^*))\}$, we get that $|a(v^*) - f_S(v^*)| \leq \mathrm{error}(v^*)$. This holds for every $v^* \in M$. $\square$

**Processing Memory.** Algorithm Hashtogram maintains (on step 2) $R \cdot T$ sums of at most $n$ bits. This requires $O(R \cdot T \cdot \log n)$ bits for processing memory.

**Runtime.** Observe that a direct implementation of (step 2 of) algorithm Hashtogram consists of summing a total of $T \approx \sqrt{n}$ bits per user, and hence results in a runtime of $\approx n^{1.5}$. As we next explain, this can be reduced to $\approx n$. First observe that for the analysis of Lemma 6.5 (specifically, for the analysis of Event $E_4$), it suffices that, for every $j \in [n]$, the entries of column $j$ of $Z$ are only pairwise independent. That is, each column of $Z$ consists of $T \approx \sqrt{n}$ pairwise independent bits. We can represent such a column using $\log T \approx \log \sqrt{n}$ bits, in which case there are at most $T \approx \sqrt{n}$ choices for the columns of $Z$ (see, e.g., Construction 3.18 in [20]). So, even though the matrix $Z$ contains $n$ columns, it has at most $T \approx \sqrt{n}$ distinct columns. Let us denote those distinct columns as $z_1, \ldots, z_T$, where $z_\gamma[t]$ denotes the bit in position $t$ in this column. We will write $Z[\cdot, j] = z_\gamma$ to indicate that the $j^{\text{th}}$ column of $Z$ is $z_\gamma$. With this notation, we can restate $a_r(t)$ (computed on step 2 of algorithm Hashtogram) as follows.

$$a_r(t) = \frac{e^\epsilon + 1}{e^\epsilon - 1} \cdot \sum_{j \in I_r} y_j \cdot Z[t, j] = \frac{e^\epsilon + 1}{e^\epsilon - 1} \cdot \sum_{1 \leq \gamma \leq T} \left( \sum_{j \in I_r \text{ s.t. } Z[\cdot, j] = z_\gamma} y_j \right) \cdot z_\gamma[t].$$

Thus, we can implement step 2 of algorithm Hashtogram by first computing $\left( \sum_{j \in I_r \text{ s.t. } Z[\cdot, j] = z_\gamma} y_j \right)$ for every $1 \leq \gamma \leq T$ and $1 \leq r \leq R$ (this amounts to summing a total of $n$ bits, and can be done it time $\approx n$). Afterwards, for every choice of $(r, t)$, computing $a_r(t)$ consists of summing $T \approx \sqrt{n}$ elements. Overall, step 2 of the algorithm can be executed in time $\approx R \cdot T \cdot T \approx n \log n$.

### 6.2.2 The Full Protocol

**Remark 6.3.** The execution of Hashtogram on step 4 is made using the parameters stated in Lemma 6.5, in order to obtain accurate answers for every fixture of $n$ queries with probability $1 - \beta$. The executions of Hashtogram on step 1 are made using the parameters stated in Lemma 6.4, in order to obtain accurate answers for every fixture of two queries with probability $255/256$. Observe that every such instantiation of Hashtogram is queried $2T$ times, and hence, some of these queries might result in inaccurate answers. Nevertheless, as will be made clear later, due to our use of error correction code, these inaccurate answers will not effect the final outcome of the algorithm.

**Lemma 6.6.** *Algorithm Bitstogram satisfies $\epsilon$-LDP.*

**Lemma 6.7.** *Let $\epsilon \leq 1$, and assume that $\log d \geq O(\log(n/\beta))$. Set $R = O(\log(1/\beta))$ and $T = O\left(\frac{\epsilon \cdot n}{\sqrt{R \log d}}\right)$. Algorithm Bitstogram returns a list $L$ of length $R \cdot T$ satisfying:*

1. *With probability $1 - \beta$, for every $(v, a) \in L$ we have that $|a - f_S(v)| \leq O\left(\frac{1}{\epsilon}\sqrt{n \log(n/\beta)}\right)$.*

2. *With probability $1 - \beta$, for every $v \in \mathcal{V}$ s.t. $f_S(v) \geq O\left(\frac{1}{\epsilon}\sqrt{n \log(d) \log(\frac{1}{\beta})}\right)$, we have that $v$ is in $L$.*

**Algorithm 8** Bitstogram

**Tool used:** Binary code (Enc, Dec), where Enc : $\mathcal{V} \to \mathcal{V}'$, correcting $\frac{1}{8}$-fraction of errors.
We use a code with constant rate, that is $\log d' = O(\log d)$, where $d = |\mathcal{V}|$ and $d' = |\mathcal{V}'|$.

**Public randomness:** Random partition of $[n]$ into $R \log d'$ subsets $I_{r,\ell}$ for $(r,\ell) \in [R] \times [\log d']$.
Random hash functions $h_1, \cdots, h_R$ mapping $\mathcal{V}'$ to $[T]$.

**Setting:** Each player $j \in [n]$ holds a value $v_j \in \mathcal{V}$. Define $S = (v_1, \cdots, v_n)$.
For every $j \in [n]$ denote $c_j = \text{Enc}(v_j) \in \mathcal{V}'$.
For $(r,\ell) \in [R] \times [\log d']$, let $S_{r,\ell} = (h_r(c_j), c_{j,\ell})_{j \in I_{r,\ell}}$, where $c_{j,\ell}$ is bit $\ell$ of $c_j$.
That is, $S_{r,\ell}$ is a database over the domain $[T] \times \{0,1\}$.

1. For $(r,\ell) \in [R] \times [\log d']$, use Hashtogram($S_{r,\ell}$) with $\frac{\epsilon}{2}$ to get $\{a_{r,\ell}(t,b) : (t,b) \in [T] \times \{0,1\}\}$.
2. For $(r,t) \in [R] \times [T]$, define $\hat{c}_{r,t} \in \mathcal{V}'$, where bit $\ell$ of $\hat{c}_{r,t}$ is $\hat{c}_{r,t,\ell} = \text{argmax}\{a_{r,\ell}(t,0), a_{r,\ell}(t,1)\}$.
3. For $(r,t) \in [R] \times [T]$, define $\hat{v}_{r,t} = \text{Dec}(\hat{c}_{r,t}) \in \mathcal{V}$.
4. Use Hashtogram($S$) with privacy parameter $\frac{\epsilon}{2}$ to obtain $a(\hat{v}_{r,t})$ for all $(r,t) \in [R] \times [T]$.
5. Return list $L = \{(\hat{v}_{r,t}, a(\hat{v}_{r,t})) : (r,t) \in [R] \times [T]\}$.

**Remark 6.4.** The assumption in Lemma 6.7 that $\log d \geq O(\log(n/\beta))$ is without loss of generality, as otherwise the universe size $d$ is small enough to allow the server to run in time linear in $d$, which makes the problem much easier. Specifically, if $d < \sqrt{n}$ then we can instantiate the frequency oracle of Lemma 6.1, and query it for every domain element. As $d < \sqrt{n}$, this can be executed in time $\approx n$ (the runtime analysis is similar to the one in Section 6.2.1). Otherwise, if $d \geq \sqrt{n}$ then we already have that $\log d \geq O(\log n)$, and (if necessary) we can pad the representation of domain elements to satisfy the assumption that $\log d \geq O(\log(n/\beta))$.

*Proof of Lemma 6.7.* Item 1 of the lemma follows directly from Lemma 6.5. We now prove item 2. Consider the following good event (over sampling $h_1, \cdots, h_R$):

**Event $E_1$ (over sampling $h_1, \cdots, h_R$):**
There exists a subset $R_1 \subseteq [R]$ of size $|R_1| \geq \frac{7}{8}R$ s.t. for every $r^* \in R_1$ and for every $v^* \in S$ satisfying $f_S(v^*) \geq \frac{n^{1.5}}{T}$ it holds that $|\{v \in S : v \neq v^* \text{ and } h_{r^*}(\text{Enc}(v)) = h_{r^*}(\text{Enc}(v^*))\}| \leq \frac{16n^{1.5}}{T}$.

We start by showing that event $E_1$ happens with high probability. To that end, fix $v^* \in S$ and fix $r^* \in [R]$. We have that

$$\mathbb{E}_{h_{r^*}}\left[|\{v \in S : v \neq v^* \text{ and } h_{r^*}(\text{Enc}(v)) = h_{r^*}(\text{Enc}(v^*))\}|\right] = \sum_{v \in S : v \neq v^*} \mathbb{E}_{h_{r^*}}\left[\mathbb{1}_{h_{r^*}(\text{Enc}(v)) = h_{r^*}(\text{Enc}(v^*))}\right] \leq \frac{n}{T}.$$

Thus, by Markov's inequality, we have that

$$\Pr_{h_{r^*}}\left[|\{v \in S : v \neq v^* \text{ and } h_{r^*}(\text{Enc}(v)) = h_{r^*}(\text{Enc}(v^*))\}| \geq \frac{16n^{1.5}}{T}\right] \leq \frac{1}{16\sqrt{n}}.$$

Assuming that $T \leq n$, there could be at most $\sqrt{n}$ "heavy" elements $v^*$ satisfying $f_S(v^*) \geq \frac{n^{1.5}}{T}$. Hence, using the union bound,

$$\Pr_{h_{r^*}}\left[\exists v^* \text{ s.t. } f_S(v^*) \geq \frac{n^{1.5}}{T} \text{ and } |\{v \in S : v \neq v^* \text{ and } h_{r^*}(\text{Enc}(v)) = h_{r^*}(\text{Enc}(v^*))\}| \geq \frac{16n^{1.5}}{T}\right] \leq \frac{1}{16}.$$

As the hash functions are independent from each other, for $R \geq 48 \ln(\frac{1}{\beta})$, by the Chernoff bound we get that with probability at least $1 - \beta$ (over sampling $h_1, \cdots, h_R$) there exists a subset $R_1 \subseteq [R]$ of size $|R_1| \geq \frac{7}{8}R$ s.t. for every

$v^*$ satisfying $f_S(v^*) \geq \frac{n^{1.5}}{T}$ and every $r^* \in R_1$ it holds that

$$|\{v \in S : v \neq v^* \text{ and } h_{r^*}(\text{Enc}(v)) = h_{r^*}(\text{Enc}(v^*))\}| \leq \frac{16n^{1.5}}{T}.$$

That is, event $E_1$ happens with probability at least $1 - \beta$. We continue the analysis assuming that event $E_1$ occurs.

---

**Event $E_2$ (over partitioning $[n]$ into $\{I_{r,\ell}\}$):**
There exists a subset $R_2 \subseteq [R]$ of size $|R_2| \geq \frac{7}{8}R$ s.t. for every $r^* \in R_2$, there exist at least $\frac{31}{32}\log d'$ choices for $\ell^* \in [\log d']$ for which $|I_{r^*,\ell^*}| \leq \frac{2n}{R \log d'}$.

---

Following Theorem 2.4 (the Poisson approximation), we analyze event $E_2$ in the Poisson case. To that end, let $\tilde{I}_{1,1}, \cdots, \tilde{I}_{R,\log d'}$ be independent Poisson random variables with mean $\frac{n}{R \log d'}$. Now fix $(r,\ell) \in [R] \times [\log d']$. Using a tail bound for the Poisson distribution (see Theorem 2.5), assuming that $n \geq 10 R \log d'$ we have that $\Pr[\tilde{I}_{r,\ell} \leq \frac{2n}{R \log d'}] \geq \frac{99}{100}$. As $\tilde{I}_{1,1}, \cdots, \tilde{I}_{R,\log d'}$ are independent, assuming that $R \log d' \geq 300 \ln(\frac{3n}{\beta})$, by the Chernoff bound we get that with probability at least $1 - \frac{\beta}{3n}$ there are at least $\frac{98}{100} R \log d'$ choices for $(r,\ell) \in [R] \times [\log d']$ s.t. $\tilde{I}_{r,\ell} \leq \frac{2n}{R \log d'}$. Hence, by Theorem 2.4, with probability at least $1 - \beta$ there are at least $\frac{98}{100} R \log d'$ choices for $(r,\ell) \in [R] \times [\log d']$ s.t. $|I_{r,\ell}| \leq \frac{2n}{R \log d'}$. If that is the case, then there must be at least $\frac{7R}{8}$ choices for $r \in [R]$ for which

$$\left|\left\{\ell \in [\log d'] : |I_{r,\ell}| \leq \frac{2n}{R \log d'}\right\}\right| \geq \frac{31}{32}\log d'.$$

That is, $\Pr[E_2] \geq 1 - \beta$.

---

**Event $E_3$ (over partitioning $[n]$ into $\{I_{r,\ell}\}$):**
For every $v^* \in S$ s.t. $f_S(v^*) \geq \frac{n^{1.5}}{T}$ there exists a subset $R_3^{v^*} \subseteq [R]$ of size $|R_3^{v^*}| \geq \frac{7}{8}R$ s.t. for every $r^* \in R_3^{v^*}$ there exist at least $\frac{31}{32}\log d'$ choices for $\ell^* \in [\log d']$ for which $|\{j \in I_{r^*,\ell^*} : v_j = v^*\}| \geq \frac{f_S(v^*)}{2R \log d'}$.

---

We analyze event $E_3$ in the Poisson case. To that end, fix $v^* \in S$ s.t. $f_S(v^*) \geq \frac{n^{1.5}}{T}$, and let $\tilde{I}_{1,1}^{v^*}, \cdots, \tilde{I}_{R,\log d'}^{v^*}$ be independent Poisson random variables with mean $\frac{f_S(v^*)}{R \log d'}$. Now fix $(r,\ell) \in [R] \times [\log d']$. Using a tail bound for the Poisson distribution (see Theorem 2.5), assuming that $\frac{n^{1.5}}{T} \geq 37 R \log d'$ we have that $\Pr[\tilde{I}_{r,\ell} \geq \frac{f_S(v^*)}{2R \log d'}] \geq \frac{99}{100}$. As $\tilde{I}_{1,1}^{v^*}, \cdots, \tilde{I}_{R,\log d'}^{v^*}$ are independent, assuming that $R \log d' \geq 300 \ln(\frac{3n^2}{\beta})$, by the Chernoff bound we get that with probability at least $1 - \frac{\beta}{3n^2}$ there are at least $\frac{98}{100} R \log d'$ choices for $(r,\ell) \in [R] \times [\log d']$ s.t. $\tilde{I}_{r,\ell} \geq \frac{f_S(v^*)}{2R \log d'}$. Hence, by Theorem 2.4, with probability at least $1 - \frac{\beta}{n}$ there are at least $\frac{98}{100} R \log d'$ choices for $(r,\ell) \in [R] \times [\log d']$ s.t. $|\{j \in I_{r^*,\ell^*} : v_j = v^*\}| \geq \frac{f_S(v^*)}{2R \log d'}$. If that is the case, then there must be at least $\frac{7R}{8}$ choices for $r \in [R]$ for which

$$\left|\left\{\ell \in [\log d'] : |\{j \in I_{r^*,\ell^*} : v_j = v^*\}| \geq \frac{f_S(v^*)}{2R \log d'}\right\}\right| \geq \frac{31}{32}\log d'.$$

Using the union bound, this holds simultaneously for every such $v^*$ with probability at least $1 - \beta$. That is, $\Pr[E_3] \geq 1 - \beta$.

---

**Event $E_4$ (over partitioning $[n]$ into $\{I_{r,\ell}\}$):**
For every $v^* \in S$ s.t. $f_S(v^*) \geq \frac{n^{1.5}}{T}$ there exists a subset $R_4^{v^*}$ of size $|R_4^{v^*}| \geq \frac{7}{8} \cdot \frac{7}{8}R$ s.t. for every $r^* \in R_4^{v^*}$ there exist at least $\frac{31}{32}\log d'$ choices for $\ell^* \in [\log d']$ for which

$$|\{j \in I_{r^*,\ell^*} : v_j \neq v^* \text{ and } h_{r^*}(\text{Enc}(v_j)) = h_{r^*}(\text{Enc}(v^*))\}| \leq \frac{32n^{1.5}}{RT \log d'}.$$

---

We analyze event $E_4$ in the Poisson case. To that end, fix $v^* \in S$ s.t. $f_S(v^*) \geq \frac{n^{1.5}}{T}$. For $r \in [R]$ denote

$$\mathrm{Col}_r(v^*) = |\{v \in S : v \neq v^* \text{ and } h_r(\mathrm{Enc}(v)) = h_r(\mathrm{Enc}(v^*))\}|,$$

and recall that by event $E_1$, for $r \in R_1$ we have that $\mathrm{Col}_r(v^*) \leq \frac{16n^{1.5}}{T}$.

Let $\tilde{I}^{v^*}_{1,1}, \cdots, \tilde{I}^{v^*}_{R,\log d'}$ be independent Poisson random variables with mean $\frac{\mathrm{Col}_r(v^*)}{R \log d'}$. Now fix $(r, \ell) \in R_1 \times [\log d']$. Using a tail bound for the Poisson distribution (see Theorem 2.5), assuming that $T \leq \frac{n^{1.5}}{R \log d'}$ we have that $\Pr[\tilde{I}^{v^*}_{r,\ell} \leq \frac{32n^{1.5}}{RT \log d'}] \geq \frac{99}{100}$. As $\tilde{I}^{v^*}_{1,1}, \cdots, \tilde{I}^{v^*}_{R,\log d'}$ are independent, assuming that $|R_1| \log d' \geq 300 \ln(\frac{3n^2}{\beta})$, by the Chernoff bound we get that with probability at least $1 - \frac{\beta}{3n^2}$ there are at least $\frac{98}{100}|R_1| \log d'$ choices for $(r, \ell) \in R_1 \times [\log d']$ s.t. $\tilde{I}^{v^*}_{r,\ell} \leq \frac{32n^{1.5}}{RT \log d'}$. Hence, by Theorem 2.4, with probability at least $1 - \frac{\beta}{n}$ there are at least $\frac{98}{100}|R_1| \log d'$ choices for $(r, \ell) \in R_1 \times [\log d']$ s.t.

$$|\{j \in I_{r^*, \ell^*} : v_j \neq v^* \text{ and } h_{r^*}(\mathrm{Enc}(v_j)) = h_{r^*}(\mathrm{Enc}(v^*))\}| \leq \frac{32n^{1.5}}{RT \log d'}.$$

If that is the case, then there must be at least $\frac{7R_1}{8}$ choices for $r \in R_1$ for which

$$\left|\left\{\ell \in [\log d'] : |\{j \in I_{r^*, \ell^*} : v_j = v^*\}| \leq \frac{32n^{1.5}}{RT \log d'}\right\}\right| \geq \frac{31}{32} \log d'.$$

Using the union bound, this holds simultaneously for every such $v^*$ with probability at least $1 - \beta$. That is, $\Pr[E_4] \geq 1 - \beta$.

We are now ready to complete the proof. Fix $v^* \in S$ s.t. $f_S(v^*) \geq \frac{264n^{1.5}}{T}$. Let $R_1, R_2, R^{v^*}_3, R^{v^*}_4$ be the sets from events $E_1, E_2, E_3, E_4$, and observe that $|R_1 \cap R_2 \cap R^{v^*}_3 \cap R^{v^*}_4| \geq \frac{3}{8}R$, and furthermore, for every $r^* \in (R_1 \cap R_2 \cap R^{v^*}_3 \cap R^{v^*}_4)$ there exist at least $\frac{29}{32} \log d'$ choices for $\ell \in [\log d']$ for which

(a) $|\{j \in I_{r^*, \ell} : v_j = v^*\}| \geq \frac{132n^{1.5}}{TR \log d'}$,

(b) $|\{j \in I_{r^*, \ell} : v_j \neq v^* \text{ and } h_{r^*}(\mathrm{Enc}(v_j)) = h_{r^*}(\mathrm{Enc}(v^*))\}| \leq \frac{32n^{1.5}}{TR \log d'}$,

(c) $|I_{r^*, \ell}| \leq \frac{2n}{R \log d'}$.

Denote $c^* = \mathrm{Enc}(v^*)$. Observe that by items (a),(b) above, we have that

$$f_{S_{r^*, \ell}}(h_{r^*}(c^*), c^*_\ell) \geq f_{S_{r^*, \ell}}(h_{r^*}(c^*), 1 - c^*_\ell) + \frac{100n^{1.5}}{TR \log d'}.$$

Let us assume that $T \leq \frac{\epsilon n}{326\sqrt{R \log d'}}$. By the properties of algorithm Hashtogram (Lemma 6.4), For every $r^*, \ell^*$ satisfying items (a),(b),(c), algorithm Hashtogram$(S_{r^*, \ell^*})$ ensures that with probability at least $\frac{255}{256}$ we have $\hat{c}_{r^*, t^*, \ell} = c^*_\ell$, where $t^* = h_{r^*}(v^*)$.

Now fix $r^* \in (R_1 \cap R_2 \cap R^{v^*}_3 \cap R^{v^*}_4)$, and recall that the coins of Hashtogram$(S_{r^*, \ell})$ are independent for different values of $\ell$. Hence, by the Chernoff bound, for $\log d' \geq 850 \log(\frac{n}{\beta})$ we get that

$$\Pr\left[|\{\ell \in [\log d'] : \hat{c}_{r^*, t^*, \ell} = c^*_\ell\}| \geq \frac{9}{10} \log d'\right] \geq 1 - \frac{\beta}{n}.$$

That is, for our fixed $v^*$ there exists an $r^* \in (R_1 \cap R_2 \cap R^{v^*}_3 \cap R^{v^*}_4)$ s.t. with probability at least $1 - \frac{\beta}{n}$ we have that $\hat{c}_{r^*, t^*}$ and $c^* = \mathrm{Enc}(v^*)$ differ on at most $\frac{1}{10}$-fraction of their bits. Using the union bound, with probability at least $1 - \beta$, this holds simultaneously for every $v^*$ s.t. $f_S(v^*) \geq \frac{264n}{T}$. By the properties of the error correction code, in such a case, for every $v^*$ s.t. $f_S(v^*) \geq \frac{264n^{1.5}}{T}$ we have that $\hat{v}_{r^*, t^*} = \mathrm{Dec}(\hat{c}_{r^*, t^*}) = v^*$, and that $v^*$ is in the list $L$. $\square$

**Runtime.** On step 1 algorithm `Bitstogram` instantiates $O(R \cdot \log d)$ copies of the frequency oracle `Hashtogram`. Every such instantiation takes time $\approx n$. Afterwards, the algorithm queries the oracles for a total of $O(2RT \log d) \approx n$ queries, each of which takes time $\tilde{O}(1)$. Thus, step 1 takes time $\approx n$ to compute. Steps 2 and 3 loop for every $(r,t) \in [R] \times [T]$, and take time $\approx RT \approx n$ to compute. Finaly, step 4 instantiates `Hashtogram` (in time $\approx n$) and queries it for a total or $R \cdot T \approx n$ queries. Overall, algorithm `Bitstogram` runs in time $\approx n$.

**Processing Memory.** Recall that step 1 of algorithm `Bitstogram` requires querying algorithm `Hashtogram` for a total of $O(2RT \log d) \approx n$ queries. As we are aiming for an algorithm with processing memory $\approx \sqrt{n}$, we cannot store all of the answers in memory simultaneously. To resolve this issue, let us reorganize step 1 of algorithm `Bitstogram` as follows:

1a. For every $(r, \ell) \in [R] \times [\log d']$, invoke $\text{Hashtogram}(S_{r,\ell})$ with $\frac{\epsilon}{2}$.

1b. For every $(r,t) \in [R] \times [T]$, for every $\ell \in [\log d']$, query $\text{Hashtogram}(S_{r,\ell})$ to get $\{a_{r,\ell}(t,b) : b\{0,1\}\}$.

Now, every one of the steps 1b,2,3,4 of the algorithm `Bitstogram` contains a loop over every choice of $(r,t) \in [R] \times [T]$, and we can group all of this steps together into one loop over $(r,t) \in [R] \times [T]$. If an iteration of this loop results in a value $a(\hat{v}_{r,t}) \leq \sqrt{n}$, we can simply ignore it (recall that the frequencies of domain elements that are not in the list $L$ are estimated as zero, and that $\sqrt{n}$ is less than the guaranteed bound on the error of the protocol, so this step does not effect our error bounds). As there could be at most $\sqrt{n}$ elements with frequencies at least $\approx \sqrt{n}$, the necessary processing memory is only $\approx \sqrt{n}$.

# A  Missing Proofs for Section 5

## A.1  Proof of Theorem 5.1

It is easy to see that each invocation of LocalRnd is $\epsilon/2$-locally differentially private since conditioned on any realization of $\ell_i, j_i, r_i$, for any pair of possible input items $v_i, v'_i \in \mathcal{V}$ to LocalRnd and any output bit $b$ generated in step 7 of Algorithm 1, we have
$$\mathbb{P}[y_i = b \mid v_i] \leq e^{\epsilon/2} \, \mathbb{P}[y_i = b \mid v'_i]$$

Note that LocalRnd is invoked only twice for each user: once when Final $= 0$ (the scanning/pruning phase of TreeHist) and another time when Final $= 1$ (the final phase of TreeHist). Thus, it follows that protocol TreeHist is $\epsilon$-differentially private.

## A.2  Proof of Theorem 5.2

Let $\beta \in (0,1)$ and $\eta$ as defined in the theorem statement. Consider the pruning phase of TreeHist, that is, Steps 1 to 11 in Algorithm 3. Let $\gamma$ be as set in Step 2. In this phase, TreeHist invokes FreqOracle once in every iteration of the outer **for** loop (over the levels of the tree) with the flag Final $= 0$. Consider any such iteration $\ell$. Suppose, for now, that the size of ChildSet(Prefixes) passed to FreqOracle in that iteration is at most $2n/\eta$. (We will show that with probability at least $1 - \beta$ this condition is satisfied at all levels $\ell$ of the tree, i.e., it's a loop invariant). By invoking Lemma 5.3 with $\widehat{\mathcal{V}} = $ ChildSet(Prefixes), $\{\tilde{\mathcal{I}}_j : j \in [t]\} = \{\mathcal{I}_{\ell,j} : j \in [t]\}$, and $\gamma = t \log d = 110 \log(n/\beta) \log d$, we have that with probability at least $1 - \beta/\log(d)$, for every $\hat{v} \in $ ChildSet(Prefixes) $: f(\hat{v}) > 3\eta$, FreqOracle gives an estimate $\hat{f}(\hat{v}) \geq 2\eta$, and for every $\hat{v} \in $ ChildSet(Prefixes) $: f(\hat{v}) \leq \eta$, FreqOracle gives an estimate $\hat{f}(\hat{v}) < 2\eta$. Hence, Step 9 implies, that with probability at least $1 - \beta/\log d$, all $\hat{v} \in $ ChildSet(Prefixes) with true frequencies $f(\hat{v}) \geq 3\eta$ will proceed to the next iteration $\ell + 1$ and all those $\hat{v} \in $ ChildSet(Prefixes) with true frequencies $f(\hat{v}) < \eta$ will be pruned out. Since the number of nodes $\hat{v}$ with true frequency $f(\hat{v}) \geq \eta$ cannot be more than $n/\eta$, then the number of surviving nodes in the next iteration $\ell + 1$ cannot be more than $2n/\eta$. Hence, this condition will be satisfied in the next iteration, and we can proceed in the same fashion. Note that when $\ell = 1$, the condition is trivially satisfied since there are only $2 < 2n/\eta$ nodes at that level. This induction argument shows that with probability at least $1 - \beta$, for every level $\ell \in [\log d]$, the surviving nodes at level $\ell$ correspond to prefixes whose true frequencies are not below $\eta$ and include all prefixes whose true frequencies are above $3\eta$. In particular, with probability at least $1 - \beta$, all items in SuccHist satisfy these properties. This covers the proof of items 1 and 2 of Theorem 5.2.

Now, consider the final phase of TreeHist, that is, Steps 12 to 14 Algorithm 3. Let $\gamma$ be as set in Step 12. In this phase, TreeHist invokes FreqOracle on the surviving nodes at the final level of the tree (the last update of Prefixes) and with input flag Final $= 1$. Now, by invoking Lemma 5.3 with $\widehat{\mathcal{V}} = $ Prefixes, $\{\tilde{\mathcal{I}}_j : j \in [t]\} = \{\mathcal{I}_j : j \in [t]\}$, and $\gamma = t = 110 \log(n/\beta)$, we have that with probability at least $1 - \beta/\log(d)$, for every $\hat{v} \in $ Prefixes, $|\hat{f}(\hat{v}) - f(\hat{v})| \leq 14\sqrt{nt}/\epsilon = O\left(\frac{\sqrt{n \log(n/\beta)}}{\epsilon}\right)$. This proves item 3 of the theorem.

# B  Missing proofs from Section 6

## B.1  Proof of Lemma 6.4

Consider the following good event:

**Event $E_1$ (over sampling $h_1, \cdots, h_R$):**
For every query $v^* \in V$ there exists a subset $R_1^{v^*} \subseteq [R]$ of size $|R_1^{v^*}| \geq \frac{7}{8}R$ s.t. for every $r^* \in R_1^{v^*}$ it holds that $|\{v \in S : v \neq v^* \text{ and } h_{r^*}(v) = h_{r^*}(v^*)\}| \leq \frac{16n}{T}$.

Event $E_1$ states that for at least $7R/8$ of the hash functions, we have that $v^*$ is mapped into a cell without too many collisions with different input elements. Informally, for every single hash function $h_r$, algorithm HashHist estimates the number of occurrences of $h_r(v^*)$ in $S$. Hence, if event $E_1$ occurs, then most of the estimations result in accurate answers. We start by showing that event $E_1$ happens with high probability. To that end, fix $v^* \in V$ and fix $r^* \in [R]$. We have that

$$\mathbb{E}_{h_{r^*}}\left[|\{x \in S : v \neq v^* \text{ and } h_{r^*}(v) = h_{r^*}(v^*)\}|\right] = \sum_{v \in S: v \neq v^*} \mathbb{E}_{h_{r^*}}\left[\mathbb{1}_{h_{r^*}(v) = h_{r^*}(v^*)}\right] \leq \frac{n}{T}.$$

Thus, by Markov's inequality, we have that

$$\Pr_{h_{r^*}}\left[|\{v \in S : v \neq v^* \text{ and } h_{r^*}(v) = h_{r^*}(v^*)\}| \geq \frac{16n}{T}\right] \leq \frac{1}{16}.$$

As the hash functions are independent from each other, for $R \geq 48\ln(\frac{d'}{\beta})$, by the Chernoff bound we get that with probability at least $1 - \beta/d'$ (over sampling $h_1, \ldots, h_R$) there exists a subset $R_1^{v^*} \subseteq [R]$ of size $|R_1^{v^*}| \geq \frac{7}{8}R$ s.t. for every $r^* \in R_1^{v^*}$ it holds that

$$|\{v \in S : v \neq v^* \text{ and } h_{r^*}(v) = h_{r^*}(v^*)\}| \leq \frac{16n}{T}.$$

Using the union bound, we have that event $E_1$ happens with probability at least $1 - \beta$. We continue the analysis assuming that event $E_1$ occurs.

Assuming that $n \geq 8R\log(2R/\beta)$, by the Chernoff bound, with probability at least $1 - \beta$ (over partitioning $[n]$ into subsets $I_1, \ldots, I_R$), for every $r \in [R]$ we have that $\frac{n}{2R} \leq |I_r| \leq \frac{2n}{R}$. We continue the analysis assuming that this is the case.

For every $r \in [R]$, let $S_r = (v_j)_{j \in I_r}$ denote a database containing the data of all users $j$ s.t. $j \in I_r$. Also for $v^* \in V$ and $r \in [R]$ denote $|S^{r,v^*}| \triangleq |\{v \in S : h_r(v) = h_r(v^*)\}|$. That is, $|S^{r,v^*}|$ is the number of users $j$ s.t. $h_r(v_j) = h_r(v^*)$. Furthermore, for $v^* \in V$ and $r \in [R]$ denote $|I_r^{v^*}| \triangleq |\{v \in S_r : h_r(v) = h_r(v^*)\}|$. That is, $|I_r^{v^*}|$ is the number of users $j$ s.t. $j \in I_r$ and $h_r(v_j) = h_r(v^*)$. Observe that $|S^{r,v^*}| \geq f_S(v^*)$ and that $|I_r^{v^*}| \geq f_{S_r}(v^*)$.

Fix $v^* \in V$. By the Chernoff bound, with probability at least $1 - \beta/d'$ (over partitioning $[n]$ into subsets $I_1, \ldots, I_R$), for every $r \in [R]$ we have that

$$\left| R \cdot |I_r^{v^*}| - |S^{r,v^*}| \right| \leq \sqrt{3R \cdot |S^{r,v^*}| \cdot \log(\frac{2Rd'}{\beta})}. \tag{4}$$

Using the union bound this holds simultaneously for every $v^* \in V$ and $r \in [R]$ with probability at least $1 - \beta$. We continue with the analysis assuming that this is the case.

**Event $E_2$ (over sampling $Z$ and the coins of the local randomizers):**
For every query $v^* \in V$ there exists a subset $R_2^{v^*} \subseteq [R]$ of size $|R_2^{v^*}| \geq \frac{7}{8}R$ s.t. for every $r^* \in R_2^{v^*}$ it holds that $\left| R \cdot a_{r^*}(h_{r^*}(v^*)) - R \cdot |I_{r^*}^{v^*}| \right| \leq \frac{e^\epsilon+1}{e^\epsilon-1} \cdot \sqrt{11nR}$.

For $v^* \in V$ and $r \in [R]$ denote $c_r(v^*) = \sum_{j \in I_r} y_j \cdot Z[h_r(v^*), j]$, and recall that algorithm Hashtogram answers the query $v^*$ with $a(v^*) = R \cdot \frac{e^\epsilon+1}{e^\epsilon-1} \cdot \text{Median}_{r \in [R]}\{c_r(v^*)\}$. Fix $v^* \in V$ and $r \in [R]$. We now analyze the expectation of $c_r(v^*)$:

$$
\begin{aligned}
\mathbb{E}[c(v^*)] &= \sum_{j \in I_r} \mathbb{E}\left[y_j \cdot Z[h_r(v^*), j]\right] \\
&= \sum_{j \in I_r:\, h_r(v_j) = h_r(v^*)} \mathbb{E}\left[y_j \cdot Z[h_r(v^*), j]\right] + \sum_{j \in I_r:\, h_r(v_j) \neq h_r(v^*)} \mathbb{E}\left[y_j \cdot Z[h_r(v^*), j]\right] \\
&= \sum_{j \in I_r:\, h_r(v_j) = h_r(v^*)} \mathbb{E}\left[y_j \cdot Z[h_r(v^*), j]\right] + \sum_{j \in I_r:\, h_r(v_j) \neq h_r(v^*)} \mathbb{E}\left[y_j\right] \cdot \mathbb{E}\left[Z[h_r(v^*), j]\right] \\
&= |\{v \in S_r : h_r(v) = h_r(v^*)\}| \cdot \frac{e^\epsilon - 1}{e^\epsilon + 1} \triangleq |I_r^{v^*}| \cdot \frac{e^\epsilon - 1}{e^\epsilon + 1}
\end{aligned}
$$

That is, $c_r(v)$ can be expressed as two sums of $\pm 1$ independent random variables: $|I_r^{v^*}|$ variables with expectation $\frac{e^\epsilon - 1}{e^\epsilon + 1}$, and $(|I_r| - |I_r^{v^*}|)$ variables with expectation $0$ (recall that $\frac{n}{2R} \leq |I_r| \leq \frac{2n}{R}$). Using the Hoeffding bound, with probability at least $43/44$ we have that $\left| c_r(v^*) - \frac{e^\epsilon - 1}{e^\epsilon + 1} \cdot |I_r^{v^*}| \right| \leq \sqrt{11n/R}$. That is,

$$
\left| R \cdot a_r(h_r(v^*)) - R \cdot |I_r^{v^*}| \right| \leq \frac{e^\epsilon + 1}{e^\epsilon - 1} \cdot \sqrt{11nR}. \tag{5}
$$

Fix $v^* \in V$, and observe that the above sums are independent for different values of $r$. Hence, using the Chernoff bound and asserting that $R \geq 132 \ln(d'/\beta)$, for that fixed $v^* \in V$, with probability at least $1 - \beta/d'$ we have that Inequality (5) holds for at least $7R/8$ choices of $r \in [R]$. Using the union bound, with probability at least $1 - \beta$, this is true for every $v^* \in V$ simultaneously. That is, event $E_2$ happens with probability at least $1 - \beta$. We continue the analysis assuming that event $E_2$ occurs. For every $v^* \in V$ we denote $R_3^{v^*} = R_1^{v^*} \cap R_2^{v^*}$.

Combining event $E_2$ with Inequality (4), we get that for every $r \in R_2^{v^*}$

$$
\left| R \cdot a_r(h_r(v^*)) - |S^{r, v^*}| \right| \leq \frac{e^\epsilon + 1}{e^\epsilon - 1} \cdot \sqrt{11nR} + \sqrt{3R \cdot |S^{r, v^*}| \cdot \log\left(\frac{2Rd'}{\beta}\right)}. \tag{6}
$$

Recall that for every $v^* \in V$ and every $r \in [R]$ we have that $|S^{r, v^*}| \geq f_S(v^*)$. Furthermore, for every $v^* \in V$ and every $r \in R_1^{v^*}$ we have that $|S^{r, v^*}| \leq f_S(v^*) + \frac{16n}{T}$. Hence, for every $v^* \in V$ and every $r \in R_3^{v^*}$ we have that

$$
\left| R \cdot a_r(h_r(v^*)) - f_S(v^*) \right| \leq \frac{e^\epsilon + 1}{e^\epsilon - 1} \cdot \sqrt{11nR} + \sqrt{3R \cdot \left(f_S(v^*) + \frac{16n}{T}\right) \cdot \log\left(\frac{2Rd'}{\beta}\right)} + \frac{16n}{T} \triangleq \text{error}(v^*).
$$

That is, for every $r \in R_3^{v^*}$ we have that $R \cdot a_r(h_r(v^*))$ is accurate up to $\text{error}(v^*)$. As $|R_3^{v^*}| \geq \frac{3}{4}R$, and as algorithm `Hashtogram` answers $v^*$ with $a(v^*)$ chosen as the median of $\{R \cdot a_r(h_r(v^*))\}$, we get that $|a(v^*) - f_S(v^*)| \leq \text{error}(v^*)$ for every $v^* \in V$.